\documentclass[accepted]{uai2026}

\usepackage[american]{babel}

\newcommand{\InputProjectFile}[2]{%
  \IfFileExists{#1.tex}{\input{#1}}{\input{#2}}%
}
\newcommand{\BibliographyProjectFile}[2]{%
  \IfFileExists{#1.bib}{\bibliography{#1}}{\bibliography{#2}}%
}

\makeatletter
\newcommand{\LoadPackageIfNotLoaded}[2][]{%
  \@ifpackageloaded{#2}{}{%
    \if\relax\detokenize{#1}\relax
      \usepackage{#2}%
    \else
      \usepackage[#1]{#2}%
    \fi
  }%
}
\makeatother

\LoadPackageIfNotLoaded[round]{natbib}
\usepackage{bm,bbm}
\usepackage{pifont}
\usepackage{adjustbox}
\usepackage{verbatim}
\usepackage{amsfonts}
\usepackage{amsmath,amssymb}
\usepackage{amsthm}
\usepackage{algorithm}
\usepackage{algorithmic}
\usepackage{mathtools}
\usepackage{thmtools} 
\usepackage{thm-restate}
\usepackage{enumitem}
\usepackage{graphicx}
\usepackage{float}
\usepackage{setspace}
\usepackage{natbib}

\usepackage{cancel}
\usepackage{tikz}
\usepackage{subcaption}
\usepackage{placeins}
\usepackage{float}
\usetikzlibrary{patterns}
\usepackage{tabularx}

\usepackage[utf8]{inputenc}

\usepackage{verbatim}

\usepackage{algorithm}

\usepackage{nicefrac} 
\usepackage{booktabs}   
\usepackage{microtype}
\usepackage{url}

\theoremstyle{plain}

\declaretheorem[name=Theorem,numberwithin=section]{theorem}
\declaretheorem[name=Lemma,numberwithin=section]{lemma}

\declaretheorem[name=Proposition,numberwithin=section]{prop}
\declaretheorem[name=Definition,numberwithin=section]{definition}
\declaretheorem[name=Example,numberwithin=section]{exmp}
\declaretheorem[name=Assumption,numberwithin=section]{assmpt}
\declaretheorem[name=Remark,numberwithin=section]{remark}

\newcommand{\eqdef}{\ensuremath{\triangleq}}

\usepackage{longtable,tcolorbox}

\newcommand\numberthis{\stepcounter{equation}{}\tag{\theequation}}

\def\cG{\mathcal{G}}

\def\cS{\mathcal{S}}

\def\mE{\mathbb{E}}

\def\mP{\mathbb{P}}

\newcommand{\game}{\cG}
\newcommand{\pol}[1]{\pi_{#1}}

\newcommand{\strat}[1]{A_{#1}}

\newcommand{\qr}[3]{\xi_{#1}^{#2}(#3)}

\newcommand{\payoff}[1]{u_{#1}}

\newcommand{\divgfunc}[1]{f_{#1}}
\newcommand{\optpol}[1]{\pi^{\ast}_{#1}}
\newcommand{\perturbgame}{\cG_{\divgfunc{}}}

\makeatletter
\@ifpackageloaded{hyperref}{%
  \hypersetup{colorlinks=true,citecolor=cyan,linkcolor=cyan,urlcolor=cyan}%
}{}
\makeatother

\allowdisplaybreaks

\title{Efficient Decentralized Learning of Generalized Quantal Response Equilibrium}

\author[1]{Zehao Zhao}
\author[2]{Apurv Shukla}
\author[1]{Rahul Jain}
\author[3]{Vijay Subramanian}

\affil[1]{Department of Electrical and Computer Engineering, University of Southern California\\
(zhaozeha, rahul.jain)@usc.edu}

\affil[2]{Sustainability Center, University of Michigan, Dearborn\\
apurvshu@umich.edu}

\affil[3]{EECS Department, University of Michigan, Ann Arbor\\
vsgsubram@umich.edu}

\begin{document}

\maketitle

\begin{abstract}
We study a solution concept for bounded rational agents in finite normal-form general-sum games called Generalized Quantal Response Equilibrium (GQRE) which generalizes Quantal Response Equilibrium~\citep{mckelvey1995quantal}. In our setup, each player can individually maximize a smooth, regularized expected utility of the mixed profiles used, reflecting both bounded rationality that subsumes stochastic choice and also individual choice of behaviors. After establishing existence under mild conditions, we present a computationally efficient no-regret decentralized learning algorithm via a smoothened version of the Frank--Wolfe algorithm. Our algorithm uses noisy gradient estimates via bandit-feedback from a simulation oracle that reports on repeated plays of the game. We analyze finite-time convergence properties of our algorithm under assumptions that ensure uniqueness of equilibrium, using a novel class of gap functions that generalize the Nash gap. We end by demonstrating the effectiveness of our method on a set of complex general-sum games such as high-rank two-player games, large action two-player games, and known examples of difficult multi-player games.
\end{abstract}

\section{Introduction}

AI agents are increasingly prevalent in many settings that involve strategic decision-making, such as in algorithmic trading or autonomous driving. Often, such systems include a variety of agents, some being rational, some sophisticated and advanced AI agents, some more limited algorithmic agents, and some being humans. A key question then is whether we can predict the outcomes of such interactions. Classical solution concepts in game theory often rely on the assumptions that all agents are fully rational, know the game parameters (or distributions), and always play best responses (or equilibrium strategies). While this assumption yields elegant theoretical constructs such as Nash Equilibrium (NE), there is plenty of evidence in behavioral game theory (see~\citep{haile2008empirical,selfridge1989adaptive,goeree2003risk}) that shows that this the assumption doesn't hold, particularly when players are faced with uncertainty, limited information, or simply cognitive constraints. As a result, alternative models have been developed to better align equilibrium predictions with observed behavior in strategic environments. Further, learning to play NE of general sum games is known to be hard in-general---impossible with deterministic dynamics for some games~\citep{milionis2023impossibility}, and also using just gradient feedback~\citep{jordan1993three,hart2003uncoupled} for others. This has been another motivation to study alternate behavioral models.

In this paper, we first introduce the Generalized Quantal Response Equilibrium (GQRE), a generalization of several existing notions of QRE~\citep{mckelvey1995quantal}. This is done by introducing a general agent-specific divergence function which can be a $\phi$-divergence, Wasserstein distance or Renyi divergence, leading to various specific notions of QRE. Thereafter, each agent chooses its response to the strategies of other players by minimizing the cost of the strategy (e.g., using an agent-specific divergence function) while preserving some minimum (expected) utility. An interpretation of this generalization is that it allows for various payoff perturbations that correspond to different risk sensitivities. Importantly, these perturbations are assumed to be chosen by the agents based on the type of bounded rationality they would like to exhibit. Empirical evidence supporting the QRE and similar payoff perturbations can be found in~\citep{goeree2003risk,selfridge1989adaptive}. Under minimal assumptions, we show existence of GQRE (Theorem~\ref{thm:properties-q-response}) by showing equivalence to a perturbed game's NE. This leads to the characterization of GQRE via a variational inequality (Lemma~\ref{lem:vi-characterization-gqre}), and a polynomial-time verification algorithm for GQRE and  approximate GQRE (Lemma~\ref{lem:verify-gqre-linear}).   

Whereas the GQRE is subsumed in the generalized equilibrium concept proposed by \citep{allen2021generalization}, our main reason to study the GQRE is to understand how agents can interact with others based on limited knowledge of the game including their payoffs, and also the behavioral choices made by other agents, but still reach some notion of equilibrium through repeated interactions. Hence, our main contribution is the analysis of a computationally efficient no-regret learning algorithm based on a smoothened Franke-Wolfe (FW) algorithm for computing the GQRE for general-sum games in the independent learning setting; the algorithm is an example of the family proposed in \citep{pena2019generalized}, where asymptotic optimality was shown with noiseless gradients. In particular, our main result establishes asymptotic and non-asymptotic performance guarantees for this algorithm (Theorem~\ref{thm:non-asymptotic-fw}) when using a simulator oracle that provides noisy bandit-feedback to the agents similar to \citep{kocak2014efficient,heliou2017learning}, which results in the noisy estimates having unbounded noise variance; we also note \citep{cui2022learning} that uses a regularized FW scheme with bandit feedback for a potential game. 
Our analysis establishes negative drift for a novel Lyapunov (gap) function; our gap function is not used by \citep{pena2019generalized}, and so it can be used to study the broader family of algorithms in the reference. The analysis of the associated variational inequality holds when the strategy space is a polytope and could be of independent interest, and it also holds for a slightly relaxed set of assumptions than in antecedent work (see Section~\ref{sec:compute-generalized-qre}). 

We validate all our results on a suite of numerical experiments including zero-sum, rank-$k$ and strongly-monotone games for strategy spaces as large as $1000$ actions---see Section~\ref{sec:numerical}. For all our instances we show that our smoothened Frank-Wolfe based independent learning algorithm outperforms existing state-of-the-art in terms of iteration complexity for a given Nash-gap, policy verification and regret. It also exhibits more stable performance relative to other schemes. Thus, in matrix games, relative to NE, GQRE incorporates features of human decision-making and is more amenable to decentralized computation.

In summary, our contributions are four-fold: (i) we study a solution concept that generalizes several existing QRE notions and allows us to study different behavioral choices of agents; (ii) we establish existence under minimal conditions and present a computationally tractable decentralized algorithm that uses bandit-feedback of payoff gradients from a simulator oracle; (iii) we prove non-asymptotic performance guarantees for the learning algorithm; and (iv) we provide numerical evidence of the algorithm’s efficacy relative to other algorithms in the literature. 

\paragraph{Related Work.} As mentioned earlier, whereas our equilibrium concept is subsumed in the concept studied in \citep{allen2021generalization}, their general concept doesn't apply to settings where agents choose their own behaviors as it needs information about the strategies of the opponents, and furthermore, they don't study computational aspects of their equilibrium. Another work that studies an equilibrium concept related to ours is \citep{melo2022uniqueness}, but the focus is on QRE and to find simple conditions that guarantee uniqueness of the equilibrium. From the computational perspective, closest to our work is the work by~\citep{gui2025statistical}---they propose related solution concepts but only ones in which the utilities are perturbed randomly via methods characterized by \citep{natarajan2009persistency}; the learning algorithm presented is exactly the dual averaging scheme of \citep{mertikopoulos2019learning}. 
The work in \citep{mazumdar2024tractable} introduces risk-averse quantal response wherein perturbations can depend on both players; their equilibrium concept is also subsumed by \citep{allen2021generalization}. Under an assumption of joint convexity of the risk-measure and with an additional QR related perturbation, \citep{mazumdar2024tractable} show that the coarse-correlated equilibrium (which are computable through no-regret schemes like follow-the-perturbed-leader) of a different game yields risk-averse quantal equilibrium; however, they don't discuss or analyze specific learning schemes. 
In contrast, we propose an efficient no-regret learning algorithm for independent learning and establish non-asymptotic (Theorem~\ref{thm:non-asymptotic-fw}) guarantees on its performance. From a computational regret characterization viewpoint, also related are the dual-averaging~\citep{mertikopoulos2019learning,hsieh2021adaptive}, optimistic gradient~\citep{golowich2020tight}, mirror-prox~\cite{nemirovski2009robust,juditsky2011solving} and proximal gradient~\cite{jordan2024adaptive} algorithms. We compare the performance of our algorithm against these in Section~\ref{sec:numerical} and demonstrate its performance advantages. 

\section{Generalized Quantal Response Equilibrium}

\textbf{Notation.}  
Consider a finite normal form game $\cG \eqdef \{N,A,\{u_{i}\}_{i \in N}\}$---$N$ is the set of players, each player $i \in N$ has a finite strategy set $\strat{i}$ and a base utility matrix~\footnote{For a given set of utility matrices across all agents, without loss of generality we can subtract the minimum utility entry, and then divide by the largest utility to reduce values to $[0,1]$ to get a strategically equivalent game.} $\payoff{i}: \prod_{j\in[N]}\strat{j} \mapsto [0,1]$. We will denote the mixed strategy profile $(\pi_1,\dotsc,\pi_N)$ with $\pi_i\in \Delta(A_i)$ for all $i\in [N]$ as $\boldsymbol{\pi}$ and also as $(\pi_{i},\pi_{-i})$ where $-i$ stands for $[N]\setminus\{i\}$. 
For agent $i\in[N]$, when convenient, we take $a_i\in A_i$ to denote a pure action and also probability distribution $\delta_{a_i}$ on $A_i$ with mass only on $a_i$. 

\textbf{Generalized quantal choice.} Following random utility models in stochastic choice theory~\citep{anderson1992discrete,hofbauer2002global}, we can also consider a random payoff model: agent $i$ selecting action $a_i$ obtains a random additional payoff $\varepsilon_i$, where $\varepsilon_i \sim f_i$ (a strictly positive density function independent of the base payoffs). Upon realization of the payoffs, the agent chooses the action with the highest realized payoff, and therefore, the probability of selecting action $a_i\in A_i$ with other players choosing $a_{-i}$ is: 
\[
\begin{split}
 \pi_i(a_i)= & p_{i}(a_i;u,a_{-i}) \eqdef \\
& \mP\left(a_i \in \arg\max_{\tilde{a}_i\in A_i}  u_{i}(\tilde{a}_i,a_{-i}) + \varepsilon_i \right).
\end{split}
\]
We will, however, consider the perspective originally proposed by~\citep{hofbauer2002global}---the goal of agent $i$ is to select a mixed strategy $\pi_i \in \Delta(\strat{i})$ with an expected payoff $\sum_{a_i\in A_i}\pi_i(a_i) \mathbb{E}_{\pi_{-i}}[u(a_i;A_{-i})]$ when the opponent(s) plays mixed strategy $\pi_{-i}$, but must pay a cost $f_{i}(\pi_i)/\lambda_i$. 
Therefore, the agent's choice given other players strategy $\pol{-i}$ arises from the Fenchel conjugate computation for $f_{i}$, namely, 
\begin{equation}
\label{eqn:qr-fixed-point}
\qr{i}{\lambda_{i}}{\pol{-i};f_{i}} \eqdef \arg\max_{\pol{i} \in \Delta(\strat{i})} 
\lambda u_i(\pi_i,\pi_{-i}) - f_i(\pol{i}),
\end{equation}
where $u_i(\pi_i,\pi_{-i})$ is shorthand for the multi-linear function $\mathbb{E}_{\prod_{i\in[N]}\pi_i}[u_i(A_1,\dotsc,A_N)]$; linearity in $\pi_i$ results in the Fenchel conjugate definition.

Before considering generalized response functions, we illustrate this for the logit Quantal Response~\citep{mckelvey1995quantal}.
The standard logit QR is given by: 
\begin{equation}\label{eqn:logitQR}
\begin{split}
&\xi_i(\pi_{-i}) \eqdef \\
&\left(\frac{\exp\left(\lambda_i \sum_{a_{-i} \in A_{-i}} \pi_{-i}(a_{-i}) u_{i}(a_{i},a'_{-i})\right)}{\sum_{a'_{i} \in A_{i}}\exp\left(\lambda_i \sum_{a_{-i} \in A_{-i}}\pi_{-i}(a'_{-i})u_{i}(a_{i},a'_{-i})\right)}\right)_{a_i \in A_i}
\end{split}
\end{equation}
can be obtained by solving \eqref{eqn:qr-fixed-point} with the specific strictly convex function $f_i(\pi_i)=\sum_{a_i\in A_i} \pi_i(a_i)\log(\pi_{i}(a_i))$. 

Via a Lagrangian relaxation, we note that \eqref{eqn:qr-fixed-point} can also be interpreted as 
\begin{align}
\begin{split}
\qr{i}{\lambda}{u,\pol{-i}}=& \arg \min_{\pi_i\in \Delta(A_i)} f_i(\pi_i)\\
& \text{s.t. } \mathbb{E}_{\pol{i}\times \prod_{j\in -i}\pol{j}}[u_i(A_i,A_{-i})] \geq \gamma_i, 
\end{split}
\label{eqn:utility_preservation}
\end{align}
where there is an equivalence\footnote{For all $\gamma_i\in \big[\min_{\pi_i\in \Delta(A_i)} \mathbb{E}_{\pol{i}\times\prod_{j\in -i}\pol{j}}[u_i(A_i,A_{-i})],$ $\mathbb{E}_{\pi_i(f_i)\times\prod_{j\in -i}\pol{j}}[u_i(A_i,A_{-i})]\big]$, we have $\lambda_i=0$ due to the form in \eqref{eqn:qr-fixed-point} using the distribution $\pi_i(f_i)$ that minimizes $f_i$ over $\Delta(A_i)$, but the discussion is still valid.} between $\gamma_i$ and $\lambda_i$ in \eqref{eqn:qr-fixed-point} since 
unconstrained choice of strategy corresponding to $\gamma_i=\min_{\pi_i\in \Delta(A_i)} \mathbb{E}_{\pol{i}\times\pol{-i}}[u_i(A_i,A_{-i})]$ and $\lambda_i=0$, and utility maximization corresponding to $\gamma_i=\max_{\pi_i\in \Delta(A_i)} \mathbb{E}_{\pol{i}\times\pol{-i}}[u_i(A_i,A_{-i})]$ and $\lambda_i \uparrow +\infty$.
We interpret the above as follows: agent $i$ chooses a mixed strategy by minimizing an agent-specific cost function $f_i(\cdot)$ subject to the (mixed) strategies preserving some expected utility based on a value given by $\gamma_i$ (equivalently, $\lambda_i$) that they choose,
leading to a generalized quantal response.
\begin{definition}[Generalized Quantal Response (GQR)]
Given a normal form game $\game$ and a strictly convex function $f_i$, the generalized quantal response of agent $i$, $\qr{}{i}{\pol{-i};f_i} \in \Delta(A_i)$, is given by:
\begin{align}
\label{eqn:generalized-q-response}
\qr{i}{\lambda_i}{\pol{-i};\divgfunc{i}} \eqdef \arg\max_{\pi_i\in \Delta(A_i)}  \lambda_{i}u_i(\pi_i,\pi_{-i}) - f_i(\pi_i).
\end{align}
\end{definition}
When the function $f_i$ measures the divergence between a base distribution, $\pi_i(f_i)$ over actions (such as uniform over all actions), the generalized quantal choice $\qr{}{i}{}$ is the distribution that minimizes the penalty paid by the player for deviating from base behavior, i.e.,
\begin{equation}
\qr{i}{\lambda_i}{\pol{-i}; \divgfunc{i}} \eqdef \arg\max_{\pol{i} \in \Delta(A_i)} \lambda_i  u_i(\pi_i,\pi_{-i}) 
- d_{f_i}(\pol{i},\pi_i(f_i)),
\end{equation}
where $d_{f_i}(\cdot,\cdot)$ is the corresponding divergence measure. Depending on the divergence measure, this generalization can incorporate different choice models such as fair response (see Example~\ref{eg:generalized-quantal}) that transcends the original motivations of best-responding with choice smoothing. 

\begin{exmp}[Divergence Measures]
\label{eg:divergence-measures}
Below we list a few divergence measures that could be used for the $f_i$ functions in a GQR---see details in Appendix~\ref{app:example_responses}. In all cases we will be considering discrete distributions, say $P$ and $Q$, over set $\{1,2,\dotsc,N\}$; we will also be taking $Q$ to be a reference distribution, say the uniform distribution $U_{\{1,2,\dotsc,N\}}$. Then, we have:\\
1. ($\phi$-divergence):
Consider a convex function $\phi$ such that $\phi(1)=0$. 
The $\phi$-divergence between two distributions $P$ and $Q$ such that $P$ is absolutely continuous with respect to $Q$ is defined as $\sum_{j=1}^n  Q(j) \phi\left(\tfrac{P(j)}{Q(j)} \right)$ (where we set $0/0=1$). Note that the KL-divergence can be obtained with $\phi(x) = x \log x$, the total variation distance can be obtained using $\phi(x) = \frac{1}{2}\vert x-1 \vert$, the $\alpha$-divergences using $\phi(x) = \frac{x^{\alpha}-1}{\alpha(\alpha-1)}$ for $\alpha \in (0,1) \cup (1,\infty)$ (with continuity used to define values of $\alpha\in\{0,1\}$), and for $alpha \geq 1$, and the $\chi^{\alpha}$-divergences using $\phi(x)=0.5 |x-1|^\alpha$.\\
2. (Wasserstein Distance): The Wasserstein distance of order $p$ between two distribution $P$ and $Q$ is given by $W_{p}(P,Q) = \inf_{\gamma \in \Gamma(P,Q)} (\mE_{(X,Y)\sim\gamma}[d(X,Y)^{p}])^{1/p}$, where $\Gamma$ is the set of all joint distributions with marginals $P$ and $Q$, and for all $x,y\in\{1,2,\dotsc,N\}$, $d(x,y)=|x-y|$.\\
3. (R\'{e}nyi Divergence): When distribution $P$ is absolutely continuous with respect to distribution $Q$, for $\alpha \in (0,1) \cup (1,+\infty)$ define $d_{\alpha}(P,Q) \eqdef \frac{1}{1-\alpha} \log \sum_{i=1}^N P(i)^{\alpha} Q^{1-\alpha}(i)$. The value at $\alpha=1$ is the KL-divergence (defined via continuity). Similarly, (via continuity), the value at $\alpha=0$ is given by $\sum_{i=1}^N Q(i) 1_{\{P(i)>0\}}$, and at $\alpha=\infty$ is given by $\log(\sup_{i\in\{1,2,\dotsc,N\}} P(i)/Q(i)$. Only for $\alpha \in [0,1]$ does convexity in $P$ hold.\\
4. (Hellinger Distance): The Hellinger distance between two distribution $P$ and $Q$ is given by $1/\sqrt{2} \sqrt{\sum_{i=1}^N (\sqrt{P(i)}-\sqrt{Q(i)})^2}$.
\end{exmp}
Convex program~\eqref{eqn:generalized-q-response} yields useful properties for a GQR.
\begin{prop}[GQR Properties]
\label{thm:properties-q-response}
Given a normal form game $\game$, for all $\lambda >0$ we have: the GQR in~\eqref{eqn:generalized-q-response} exists if $f_i$ is convex,
and is unique and continuous in $\pol{-i}$ if $f_i$ is strictly convex
\end{prop}

\begin{exmp}
\label{eg:generalized-quantal}
Consider the GQR for a divergence measure from Example \ref{eg:divergence-measures}---see details in Appendix~\ref{app:example_responses}.\\
(Fair Quantal Response)
Consider the total variation distance---set $f(t) = \frac{1}{2}\vert t- 1\vert$) in~\eqref{eqn:generalized-q-response}. This leads to a sparse response function where actions with low utility may get probability zero, in contrast to the KL-divergence regularization in~\eqref{eqn:logitQR}. (Also, see \citep[Example 4.10]{gui2025statistical}.)
\end{exmp}


\begin{remark} 
\label{rem:gqr-taskesen}
We emphasize the following: 
(i) Classical models of fictitious play~\cite{fudenberg1998theory,hofbauer2002global} and games with penalized cost~\cite{van1982refining}, consider penalty functions which are infinitely steep at the boundary of the simplex with positive-definite Hessian. In this paper, the functions $f_i$ need only be strictly convex (or even just convex for existence).\\
(ii) The GQR  in~\eqref{eqn:generalized-q-response} is not a regular quantal response as per~\citep{goeree2005regular} since our choice functions allow for zero mass over some actions (see Example~\ref{eg:generalized-quantal}), and thus, do not satisfy their interior criteria.\\
\end{remark}

Let $\mathcal{P}(\cS)$ denote the power set given the set $\cS$. Then, we have the following definition.
\begin{definition}[$(\boldsymbol{\lambda}, \boldsymbol{f})$-Generalized Quantal Response Equilibrium (GQRE)]
\label{defn:generalized-qre}
Define the correspondence $\Gamma^{(\boldsymbol{\lambda}, \boldsymbol{f})}_{\text{GQRE}}:\prod_{i\in[N]}\Delta(A_i)\mapsto \prod_{i\in[N]}\mathcal{P}(\Delta(A_i))$ as: 
\[
\Gamma^{(\boldsymbol{\lambda}, \boldsymbol{f})}_{\text{GQRE}}(\boldsymbol{\pi})\eqdef \big\{(\tilde{\pi}_i,\dotsc,\tilde{\pi}_N): \forall \tilde{\pi}_i\in \xi_i^{\lambda_i}(\pi_{-i};f_i)\ \forall i\in [N]\big\}.
\] 
Then, a GQRE is given by an $N$-tuple of policies $\boldsymbol{\pi}^*=(\pi_1^*,\dotsc,\pi_N^*)$ that satisfy
$$\boldsymbol{\pi}^*\in \Gamma^{(\boldsymbol{\lambda}, \boldsymbol{f})}_{\text{GQRE}}(\boldsymbol{\pi}^*), \text{ i.e., } \pi_i^*\in \xi_i^{\lambda_i}(\pi_{-i}^*;f_i),\ \forall i\in[N].$$
\end{definition}

We will also consider the notion of an approximate equilibrium concept, $\epsilon$-GQRE. 
\begin{definition}[$(\boldsymbol{\lambda}, \boldsymbol{f})$ $\epsilon$-GQRE]
\label{defn:epsilon-gqre}
For $\epsilon\geq 0$, a strategy profile $\boldsymbol{\pi}^*$ is a $(\boldsymbol{\lambda}, \boldsymbol{f})$ $\epsilon$-GQRE if 
\begin{align*}
    & ~~\forall i \in [N],\ \lambda_i u_i(\pi_i^*,\pi_{-i}^*)- f_i(\pi_i^*)  \geq  \\
    & \lambda_i u_i\left(\xi_i^{\lambda_i}(\pi_{-i}^*;f_i),\pi_{-i}^*\right) - f_i\left(\xi_i^{\lambda_i}(\pi_{-i}^*;f_i)\right)-\epsilon, 
\end{align*}
i.e., the utility achieved is within $\epsilon$ of the best possible utility for every agent given the strategy profile.
\end{definition}





Definition~\ref{defn:generalized-qre} defines a GQRE as a fixed point a'la Nash Equilibrium (NE). We now show that a GQRE for a game $\game$ is an NE for a perturbed game $\perturbgame$. Given game $\game= \{[N],\{\strat{i}\},\{u_i\}_{i \in N}\}$ and functions $\{f_{i}\}_{i \in [N]}$, for a mixed strategy profile $\boldsymbol{\pi}=(\pi_1,\pi_2)$, define the perturbed utilities as 
$$u^{f_i}_{i}(\pol{i},\pol{-i})\eqdef\lambda_{i} u_i(\pol{i},\pol{-i}) - f_{i}(\pol{i}),$$ 
and the perturbed game $\perturbgame$ as $\perturbgame = \{[N],\{\Delta(A_i)\}_{i \in N}, \{u^{f_i}_{i}\}_{i \in [N]}\}$. 

\begin{assmpt}
\label{assmpt:lipschitz-continuity} We assume the following for the perturbed utilities: 
(i) The gradients of the perturbed utilities are Lipschitz-continuous, and 
(ii) The gradients of the perturbed utilities have a Lipschitz-continuous Jacobian matrix $H$ with each element defined as
\[ 
H_{(i,a_i),(j,a_j)}(\boldsymbol{\pi}) = \frac{\partial^2 u_i^{f_i}(\boldsymbol{\pi})}{\partial \pi_{i,a_i} \partial \pi_{j,a_j}},\, \forall i, j \in [N], a_i \in A_i, a_j \in A_j
\]
\end{assmpt}

\begin{assmpt}[(Strict) Diagonal Dominance]
\label{assmpt:diagonal-dominance}
The game $\perturbgame$ satisfies strict diagonal dominance if the matrix $H(\boldsymbol{\pi})+H^T(\boldsymbol{\pi})$ is negative definite for all $\boldsymbol{\pi}\in \prod_{i\in[N]} \Delta(A_i)$.
\end{assmpt}
Since $u_i^{f_i}(\boldsymbol{\pi})=\lambda_i u_i(\pi_i,\pi_{-i}) - f_i(\pi_i)$ for each $i\in [N]$, $H$ has a block-form where the $H_{(i,\cdot),(j,\cdot)}$ block depends only on $f_i$, if $i=j$ (which in fact, is the Hessian $\Delta f_i$) and $u_i$ the utility matrix of agent $i$. 
We will use this important property in our analysis.
With $\{f_i\}_{i\in[N]}$ being convex, we have a concave game, and by \citep{rosen1965existence} an NE always exists. Also, Assumption~\ref{assmpt:diagonal-dominance} is exactly the condition for a unique NE from \citep{rosen1965existence}. We will also assume that the unique equilibrium is fully mixed. We summarize this next.


\begin{theorem}[GQRE of $\game$ is NE of $\perturbgame$]
\label{thm:gqre-ne}
A pair of policies $(\pi_1^*,\pi_2^*)$ is the GQRE for $\game$ if and only if they are the Nash Equilibria for $\perturbgame$. Since $\perturbgame$ is a concave game when $\{f_i\}_{i\in[N]}$ are convex functions, a GQRE always exists. Further, by Assumption~\ref{assmpt:diagonal-dominance} of strict diagonal dominance, the GQRE is unique.
\end{theorem}
Under a bounded influence condition, \cite{melo2022uniqueness} studied Theorem~\ref{thm:properties-q-response} for QRE and developed simple conditions for uniqueness.
\cite{gui2025statistical,mertikopoulos2019learning} use \citep{rosen1965existence} as well; again, concept of \cite{allen2021generalization} also fits. 

\begin{remark}\label{eq:gqre_exploration}
    In our definition of GQR and GQRE, we assumed that each agent $i$ chose its (mixed) strategy from $\Delta(A_i)$. We can also assume that each agent $i$ chooses its strategy from the $\epsilon_i$-exploration simplex $\Delta(A_i,\epsilon_i)$ where the probability of choosing each action is at least $\epsilon_i$ for $\epsilon_i \in [0, \tfrac{1}{|A_i}]$. The existence and uniqueness of an $\boldsymbol{\epsilon}$-exploration GQRE follow by the same logic as for a GQRE. We note that learning such an equilibrium will be covered naturally by our algorithm.
\end{remark}



An immediate consequence of Theorem~\ref{thm:gqre-ne} is a variational inequality  \citep{facchinei2003finite} characterizationfor GQRE as an alternative to the fixed-point characterization in Definition~\ref{defn:generalized-qre}. 
\begin{lemma}[VI characterization]
\label{lem:vi-characterization-gqre}
GQREs of the game $\game$ are solutions to variational inequality below: 
\begin{equation}
\label{eqn:vi-character}
\pi^{\ast} \ \text{s.t.} \ \langle \pi-\pi^{\ast}_{i}, \nabla_{\pi} u^{f_i}_{i}(\boldsymbol{\pi}^*) \rangle \le 0, \ \forall \ \pi \in \Delta(A_i), \forall i \in [N]
\end{equation} 
Further, \eqref{eqn:vi-character} is monotone since $u_i^{f_i}$ is concave in $\pi_i$, and strongly monotone if Assumption~\ref{assmpt:diagonal-dominance} holds.
\end{lemma}

The variational inequality characterization helps in designing efficient computational methods for computing the generalized quantal response equilibria by borrowing from the rich literature on optimization and variational inequalities. It also leads to a verification procedure for checking whether a given strategy profile is a GQRE or an $\epsilon$-GQRE. 

\textbf{GQRE Verification.}
Having established a notion of GQRE, we propose an algorithmic verification mechanism to check whether a given strategy $(\pi_{i}, \pi_{-i})$ is a GQRE. For this, we use Theorem \ref{thm:gqre-ne}, i.e., the equivalence between GQRE of $\game$ and NE of $\perturbgame$. Checking whether a strategy is an NE for $\perturbgame$ is equivalent to checking if $\pol{i}$ is in the best-response set of player $i$.
Since the utility functions are concave, this will be a convex optimization problem. Consequently, performing this check for all players can be done in polynomial-time. 

\begin{lemma}[Verification of GQRE and $\epsilon$-GQRE]
\label{lem:verify-gqre-linear}
The following hold:\\
1. A strategy profile $\boldsymbol{\pol{}}^*$ is a GQRE if and only if 
\begin{align*}
    \max_{a_i\in A_i} \langle \delta_{a_i} - \pi_i^*, \nabla_{\pi_i} u_i^{f_i}(\pi^*) \rangle \leq 0,~\forall i\in [N].
\end{align*}
This can be verified in $O(\vert N \vert \vert A\vert)$ time.\\
2. A given strategy $(\pol{i},\pol{-i})$ is an  $\epsilon$-GQRE for
\begin{align*}
    & \epsilon=\max_{i\in[N]} \epsilon_i, ~\text{where}\\ 
    &\epsilon_i = u_i^{f_i}(\xi_i^{\lambda_i}(\pol{-i};f_i),\pol{-i}) - u_i^{f_i}(\pol{i},\pol{-i}), ~\forall i\in[N],
\end{align*}
and determining the Nash gap $\epsilon$ requires solving $N$ convex programs.
\end{lemma}

\section{Computing a GQRE}
\label{sec:compute-generalized-qre}

Having discussed the concept of a GQRE and its existence, we now discuss a means to compute it. From Lemma~\ref{lem:vi-characterization-gqre} it is clear that gradient schemes can be an option if the exact gradient or a noisy version of it is available. This has been a fruitful avenue in the optimization literature---using the exact gradient via the prox method as in \cite{nemirovski2004prox}, or using a noisy gradient given by a martingale difference sequence (MDS) as in \cite{nemirovski2009robust,juditsky2011solving,jordan2024adaptive,mertikopoulos2019learning,heliou2017learning,golowich2020tight,hsieh2021adaptive}. 
Compared to the bulk of the literature that uses bounded conditional variance of the gradient estimates, we will use bandit feedback from a simulator oracle as in \cite{heliou2017learning,kocak2014efficient}, for which we neither have bounded variance nor independence across agents.

\textbf{Gradient access via bandit feedback.} 
Our simulator based oracle takes (mixed) strategies from all the agents, and plays the underlying finite game $M\geq 1$ times (independently if mixed strategies are used). The oracle then sends agent $i$ a table listing $M_i$, the number of times action $a_i\in A_i$ was played, and the cumulative payoff $U_i(a_i)$ for the action over the $M$ plays. Then, given strategy profile $\boldsymbol{\pi}$ and $M$ plays $\{\boldsymbol{A}(m)=(A_1(m),\dotsc,A_N(m))\}_{m=1}^M$ \emph{i.i.d.} with distribution $\prod_{i\in[N]} \pi_i$, by taking expectations (given $\boldsymbol{\pi}$)  
\begin{align}\label{eqn:utility_estimate}
\begin{split}
    & U_i(a_i) = \sum_{m=1}^M u_i(\boldsymbol{A}(m)) 1_{\{A_i(m)=a_i\}},\text{ and} \\
    & \mE[U_i(a_i)] = M \pi_{i}(a_i) u_i(a_i,\pi_{-i}).
\end{split}
\end{align}
Hence, assuming that $\pi_i(a_i)>0$, an unbiased estimate of the gradient of agent $i$'s perturbed utility from just the finite game is given $\forall a_i \in A_i$ by
\begin{align}\label{eqn:gradient_estimate}
\begin{split}
    &\hat{F}_i(a_i) =  \frac{\lambda_i}{\pi_i(a_i)} \frac{U(a_i)}{M} \text{ with}\\ &\mathrm{Var}(\hat{F}_i(a_i)) = \frac{\lambda_i^2\widetilde{\mathrm{Var}}(a_i,\pi_i,\pi_{-i};u_i)}{M \pi_i(a_i)}, 
\end{split}
\end{align}
where we have $\forall a_i \in A_i$,
\begin{align}\label{eq:varterm}
\begin{split}
& \widetilde{\mathrm{Var}}(a_i,\pi_i,\pi_{-i};u_i)\eqdef\\
& \sum_{\substack{\boldsymbol{a}_{-i}\in \\ \prod_{j\in -i} A_j}} \pi_{-i}(\boldsymbol{a}_{-i}) u_i^2(a_i,\boldsymbol{a}_{-i}) - \pi_i(a_i) u_i^2(a_i,\pi_{-i}).
\end{split}
\end{align}
Note that $\forall a_i \in A_i$, $\mathrm{Var}(\hat{F}_i(a_i))$ is unbounded over the range $\pi_i(a_i) \in (0, 1]$; this complicates algorithm design and analysis. Note that due to independent plays, the correlations of the gradient estimates across each agent $i$'s actions are easily computed. Finally, the estimates being obtained from a common dataset are dependent across agents. Since each agent $i$ knows its $\pi_i$ and $f_i$, it can then form an unbiased estimate $F_i=\hat{F}_i - \nabla_{\pi_i} f_i(\pi_i)$ of $\nabla_{\pi_i} u_i^{f_i}(\boldsymbol{\pi})$. The variance and distributional characterizations still hold unchanged.


 

\textbf{Algorithm Design.}
Another departure from existing work is in the choice of algorithmic schemes used by us to compute GQRE. Most existing work focuses on schemes such as dual-averaging, projected gradient descent, optimistic gradient descent, or extra-gradient methods to compute the solution to carefully chosen optimization problem(s). 
We instead focus on a projection-free method---the Frank-Wolfe algorithm. However, the Frank-Wolfe scheme is hard to analyze as the choice of directions suggested can be set-valued---assuming current utility gradient $\nabla_{\pi_i} u_i^{f_i}(\boldsymbol{\pi})$ for agent $i\in[N]$, in the Frank-Wolfe scheme, the agent chooses probability vector $s_i^*$ such that
\begin{align*}
    s_i^* \in \Gamma_i^{\mathrm{FW}}(\boldsymbol{\pi})\eqdef\arg\max_{s\in \Delta(A_i)} \langle s, \nabla_{\pi_i} u_i^{f_i}(\boldsymbol{\pi}) \rangle,
\end{align*}
and then updates its strategy to $\pi_i + \gamma (s_i^* - \pi_i)$ with step-size $\gamma\in (0,1)$. Whereas the VI characterization shows that $\pi_i^* \in \Gamma^{\mathrm{FW}}(\boldsymbol{\pi}^*)$ for all $i\in[N]$ holds if and only if $\boldsymbol{\pi}^*$ is GQRE, and the performance of this algorithm is good (see Section~\ref{subsec:results}), it is hard to analyze. The main reason for this is that the gap function~\footnote{Gap function, $\sum_{i\in[N]} \max_{s\in \Delta(A_i)} \langle s-\pi_i, \nabla_{\pi_i} u_i^{f_i}(\boldsymbol{\pi}) \rangle$, is differentiable only if the maximizer is unique.} for this algorithm (a potential Lyapunov function) is Lipschitz but not everywhere differentiable. Due to this issue, we smoothen the Frank-Wolfe algorithm vis a ``proximal" perturbation: $\forall i\in[N],$
\begin{align}\label{eqn:sFW0}
\begin{split}
    &  s_i^*(\boldsymbol{\pi})  \eqdef \arg \max_{s \in \Delta(A_i)} \langle s, \nabla_{\pi_i} u_i^{f_i}(\boldsymbol{\pi})\rangle - \eta \mathrm{KL}(s,\pi_i), \\
    & \forall a_i \in A_i,\ s_{i,a_i}^*(\boldsymbol{\pi})  = \frac{\pi_{i,a_i}\exp\left(\frac{1}{\eta}\nabla_{\pi_i} u_{i,a_i}^{f_i}(\boldsymbol{\pi})\right) }{\left\langle\pi_i,\exp\left(\frac{1}{\eta}\nabla_{\pi_i} u_i^{f_i}(\boldsymbol{\pi})\right) \right\rangle},
\end{split}
\end{align}
and thereafter, we define $\forall i\in[N]$,
\begin{align}\label{eqn:newgapi}
\begin{split}
    {V}_i(\boldsymbol{\pi}) & \eqdef \max_{s \in \Delta(A_i)} \langle s - \pi_i, \nabla_{\pi_i} u_i^{f_i}(\boldsymbol{\pi})\rangle - \eta \mathrm{KL}(s,\pi_i) \\
    & = \eta \log\left(\left\langle \pi_i, \exp\left(\frac{1}{\eta}\nabla_{\pi_i} u_i^{f_i}(\boldsymbol{\pi})\right)\right\rangle \right) \\
    &\quad  - \eta \left\langle \pi_i, \frac{1}{\eta} \nabla_{\pi_i} u_i^{f_i}(\boldsymbol{\pi})\right\rangle, 
\end{split}\\
    \label{eqn:newgap} \text{and set } & {V}(\boldsymbol{\pi})  \eqdef \sum_ {i\in [N]} \bar{V}_i(\boldsymbol{\pi}),
\end{align}
where we use the Donsker-Varadhan variational formula to solve for the optimizer and obtain the optimum value. Since choosing $\tilde{\pi}=\pi_i$ makes the objective in the optimization defining ${V}_i$ to take value $0$, the value of ${V}_i(\boldsymbol{\pi})$ is non-negative for each $\pi_i\in\Delta(A_i)$; this also holds using Jensen's inequality. Further, by the VI characterization for a GQRE $\boldsymbol{\pi}^*$, each ${V}_i(\boldsymbol{\pi}^*)=0$ since at a non-GQRE $\boldsymbol{s}$,
the linear term is non-positive by the VI, and the KL-divergence term is negative. Further, for a non-GQRE $\boldsymbol{\pi}$, starting from $\tilde{\pi}=\pi_i$, we can make a small enough perturbation in the right direction so that objective is positive; for this we use the fact that gradient of $\mathrm{KL}(\tilde{\pi}|\pi_i)$ is $0$ at $\tilde{\pi}=\pi_i$. This proves that $\bar{V}_i(\boldsymbol{\pi})$ is a valid smooth gap function that we can use to study convergence to GQRE---see Appendix~\ref{app:gapfunction} for details. However, the variance of the gradient estimate being unbounded discussed earlier coupled with the unbounded gradient of the KL divergence in the second term, forces us to solve for the feasible direction within the $\epsilon$-exploration simplex $\Delta(A_i;\epsilon)$ for $\epsilon \in (0,\tfrac{1}{|A_i|}]$, i.e., solving for
\begin{align}\label{eqn:sFWeps}
    \forall i\in[N],\ s_i^*(\boldsymbol{\pi};\epsilon) & \eqdef \arg \max_{s \in \Delta(A_i;\epsilon)} \frac{1}{2} \|s- s_i^*(\boldsymbol{\pi})\|_2^2 .
\end{align}
Using the unique $\lambda^*\in[\epsilon-\max_{a_i\in A_i} s^*_{i,a_i}(\boldsymbol{\pi}), 0]$ that solves (using a line search) 
\begin{align}\label{eqn:lambdasol}
    \sum_{a_i \in A_i} \max\left(s^*_{i,a_i}(\boldsymbol{\pi}) -\epsilon+\lambda,0\right) = 1-|A_i| \epsilon,
\end{align}
we set $\forall a_i\in A_i$,
\begin{align}\label{eqn:sol_sFWeps}
\begin{split}
    s_{i,a_i}^*(\boldsymbol{\pi};\epsilon)&=\epsilon+\max\left(s^*_{i,a_i}(\boldsymbol{\pi}) -\epsilon+\lambda^*,0\right) \\
    &=\max(s^*_{i,a_i}(\boldsymbol{\pi})+\lambda^*,\epsilon).
\end{split}
\end{align}
\begin{algorithm}[bthp]
\caption{Smoothened Frank-Wolfe for decentralized learning with bandit feedback}
\label{algo:fw-decentralized}
\begin{algorithmic}[1]
\STATE{\parbox[t]{0.93\linewidth}{Initialise: all players $i\in [N]$ choose a policy $\pol{i,0}$ and parameter $\eta>0$; choose sequences $\{(\gamma_t,\epsilon_t)\}_{t=1}^T$ such that, for all $t\in [T]$, $\gamma_t\in(0,1)$, $\epsilon_t \in (0,\tfrac{1}{N}]$, and $\epsilon_{t+1}\leq \epsilon_t$.}}
\FOR{$t=1:T$}
\STATE{All players $i\in [N]$ submit their policy $\pol{i,t-1}$ to the oracle}
\STATE{\parbox[t]{0.93\linewidth}{Oracle plays the game $M$ times with policy profile $\boldsymbol{\pi}_{t-1}=(\pi_{1,t-1},\dotsc,\pi_{N,t-1})$: $\{\boldsymbol{A}_t(m)=(A_{1,t}(m),\dotsc,A_{N,t}(m))\}_{m=1}^M$ are \emph{i.i.d.} with distribution $\prod_{i\in[N]}\pi_{i,t-1}$.}}
\STATE{\parbox[t]{0.93\linewidth}{Oracle returns outcomes: for $i\in [N]$ and all $a_i\in A_i$, $M_t(a_i)=\sum_{m=1}^M 1_{\{A_{i,t}(m)=a_i\}}$ and $U_t(a_i)=\sum_{m=1}^M u_i(\boldsymbol{A}_t(m))1_{\{A_{i,t}(m)=a_i\}}$.}}
\FOR{$i \in [N]$}
\STATE{Generate payoff gradient from game play: for all $a_i\in A_i$, $\hat{F}_{i,t}(a_i)=\frac{\lambda_i}{\pi_{i,t-1}(a_i)}\frac{U_t(a_i)}{M}$}
\STATE{Generate noisy payoff gradient at $\boldsymbol{\pi}_{t-1}$: $F_{i,t}=\hat{F}_{i,t}-\nabla_{\pi} f_i(\pi)|_{\pi=\pi_{i,t-1}}$}
\STATE{$\tilde{s}_{i,t} \leftarrow \arg\max_{s \in \Delta(A_{i})} \langle s, F_{i,t}\rangle - \eta \mathrm{KL}(s|\pi_{i,t-1})$}
\STATE{$s_{i,t} \leftarrow \arg\max_{s \in \Delta(A_{i};\epsilon_t) } \frac{1}{2} \|s- \tilde{s}_{i,t}\|_2^2$}
\STATE{$\pi_{i,t} \leftarrow \pi_{i,t-1} + \gamma_t \left(s_{i,t}-\pol{i,t-1}\right)$} 
\ENDFOR
\ENDFOR
\end{algorithmic}
\end{algorithm}
Thus, the algorithm has all agents independently updating their strategy via their part of $\boldsymbol{\pi}+\gamma(\boldsymbol{s}^*(\boldsymbol{\pi};\epsilon)-\boldsymbol{\pi})$. This, then, leads to the independent and decentralized learning scheme outlined in Algorithm~\ref{algo:fw-decentralized}. As mentioned earlier, the basic algorithm (without adjusting for unbounded variance) is within the family of algorithms in~\citep{pena2019generalized}, and method used to counter the unbounded variance is different from \citep{heliou2017learning} (convex combination with the uniform distribution) and \citep{kocak2014efficient} (adding a constant to the denominator); also see \citep{cui2022learning} where a positive definite matrix is used for a Frank-Wolfe scheme. We note that our main result Theorem~\ref{thm:non-asymptotic-fw} also applies to the schemes used by \citep{kocak2014efficient,heliou2017learning}, but from results in Section~\ref{sec:numerical} we will also see that our choice provides a distinct advantage. 

\textbf{Smoothened Gap.} Since we have a gap function $V(\boldsymbol{\pi})$ which is non-negative and $0$ exactly for a GQRE, we assess the performance of Algorithm~\ref{algo:fw-decentralized} using the expected gap value at the $T^{\mathrm{th}}$ timestep given by $R(T)=\mE[V(\boldsymbol{\pi}_{T})]$, 
where the randomness is induced by bandit feedback based gradient estimates.

\begin{theorem}[No-regret guarantee]
\label{thm:non-asymptotic-fw}
Under Assumptions~\ref{assmpt:lipschitz-continuity} and~\ref{assmpt:diagonal-dominance}, and at time-step $t$ step-size $\gamma_t=\tfrac{1}{t+1}$, $\epsilon_t=\tfrac{1}{(t+1)\max_{i}|A_i|}$ and the number of samples $M$ from the simulator chosen as $M_t=\lceil \tfrac{1}{\epsilon_t \gamma_t^2}\rceil$, the expected gap value of Algorithm~\ref{algo:fw-decentralized} at time $T\geq 1$ satisfies: 
\[
R(T) \le \frac{C_7}{T} + \frac{C_{6}\log(T)}{T},
\]
and further, $\lim_{T\rightarrow\infty} V(\boldsymbol{\pi}(T))=0$ in probability.
\end{theorem}

\begin{proof}
The proof in Appendix~\ref{app:proof} follows by showing that $V(\boldsymbol{\pi})$ is a valid Lyapunov function by proving that it has a negative drift. We carefully account for the noise due to the simulator, and here the choice of the number of samples is crucial. The fact that for all $i\in[N]$, $\pi_{i,t} \in \Delta(A_i,\epsilon_t)$ with $\epsilon_t>0$, so that $\pi_{i,t}(a_i) \geq \epsilon_t > 0$ for $a_i\in A_i$, is also crucial along with $\epsilon_{t+1} \leq \epsilon_{t}$; see line 1 of Algorithm~\ref{algo:fw-decentralized}. The Fenchel conjugate representation in the algorithm and Fenchel duality are also critical.
\end{proof}
The proof ideas also extend the result to the schemes in \citep{kocak2014efficient,heliou2017learning}---see Appendix~\ref{app:proof}.


\section{Numerical Results}
\label{sec:numerical}

\textbf{Games:} We evaluate \textbf{Algorithm~1} on two families of two-player games: 
(1) \textit{Strongly Monotone Games} of size $n \times n$ with $n \in \{10, 20, 100, 200\}$; and (2) \textit{Rank-$k$ General-Sum Games} of size $m \times m$ with $m = 5$ and rank parameter $k \in \{1, 2, 3, 4\}$. All experiments use the same hyperparameters: $\mu = 1.0$, $\mathsf{skew} = 0.3$ (used only in the strongly monotone setting), $T = 1000$ iterations, $M = 100$ samples per gradient estimate, step size $\eta = 1.0$, $\lambda_i = 1.0$, and $f_i(\pi_i) = \sum_{a_i\in A_i} \pi_{i,a_i} \,\log p_{i,a_i}$ for mixed strategy $\pi_i$; we solve for logit QRE; results for other penalty functions are presented in the Appendix. Each result is averaged over 20 independent runs. At each iteration, we draw $M$ simulator samples to form stochastic gradient estimates and evaluate convergence by tracking the GQRE gap over time.



\noindent\textit{Strongly Monotone Games:}
For $n\in\{10, 20, 100, 200\}$, we generate $S = \mu\,I_n$, ~$M \sim \mathcal{N}(0,1)^{n\times n}$, 
$K = \tfrac{M - M^\top}{2}$, $K \leftarrow \tfrac{\mathsf{skew}}{\|K\|_2}\,K$, and 
$A = S + K,\quad B = S - K$,
with $\mu=1.0$, $\mathsf{skew}=0.3$. This construction ensures strong monotonicity by making the sum $A + B = 2\mu I_n$ positive definite. The symmetric component $S = \mu I_n$ guarantees strong convexity–concavity, while the antisymmetric matrix $K$ introduces structured interaction between the players without violating monotonicity. The resulting game models have a unique and well-behaved NE.

\noindent\textit{Rank‑$k$ General‑Sum Games:}
For $m\times m$ games with $m=5$ and each $k\in\{1,2,3,4\}$, we sample
\begin{align*}
& U,V\sim \mathcal{N}(0,1)^{5\times k},\ 
M = U\,V^\top,\\
& K \sim \mathcal{N}(0,1)^{5\times5},\ 
S = \tfrac12\,(K - K^\top),
\end{align*}
and set $(A,B)=(M+S,\;M-S)$ so that $\mathrm{rank}(A+B)=k$. Here, $M$ is a low-rank matrix of rank at most $k$ that controls the cooperative component of the game, while $S$ introduces an antisymmetric term that governs the competitive interaction. By construction, the sum $A + B = 2M$ has rank $k$, ensuring that the underlying game lies on a low-dimensional manifold. These settings are useful for evaluating algorithm performance in structured general-sum environments with varying complexity \cite{kannan2010games}.

\noindent\textbf{The Baseline Algorithms:} We consider the following state-of-the-art algorithms as baselines for comparison: (i) \textit{Uniform-Mix Smoothened Frank-Wolfe:} Algorithm~\ref{algo:fw-decentralized} with the \citep{heliou2017learning} unbounded variance combating scheme instead of Step 10;  
(ii) \textit{Bandit Frank–Wolfe (Hard FW)} \citep{frank1956algorithm}: replaces the smoothed best response with a hard $\arg\max$ on the stochastic gradient estimate within the Frank–Wolfe update, isolating the effect of smoothing; 
(iii) \textit{Extragradient (EG)} \citep{chen2022sample}: performs a Mirror–Prox update by taking a provisional gradient step followed by a corrective gradient step, yielding robust convergence in saddle‐point problems; 
(iv) \textit{Optimistic Gradient Descent (OGD)} \citep{jordan2024adaptive}: 
uses the update \(\log \pi\leftarrow\log \pi+\gamma_t(2G_t-G_{t-1})\) to add a “predictor–corrector” term that accelerates convergence under predictable gradient sequences; and 
(v) \textit{Adaptive OGD–PGD (PGD)} \citep{jordan2024adaptive}: applies an exponentiated gradient step with \(\gamma_t=1/\sqrt t\) and then projects onto an \(\varepsilon\)-simplex to adaptively balance exploration and exploitation under stochastic noise. We do not present dual-averaging schemes from \citep{mertikopoulos2019learning,hsieh2021adaptive} as their performance was significantly worse.

\subsection{The  Results}\label{subsec:results}


\begin{table*}[t!]
\centering
\begin{tabular}{lccccc}
\toprule
 & action\_0 & action\_1 & action\_2 & action\_3 & action\_4 \\
\midrule
Entropy--Entropy & 0.183 & 0.184 & 0.149 & 0.302 & 0.182 \\
Entropy--TV      & 0.189 & 0.192 & 0.171 & 0.262 & 0.186 \\
L2--Entropy      & 0.169 & 0.171 & 0.130 & 0.363 & 0.167 \\
L2--TV           & 0.179 & 0.185 & 0.155 & 0.307 & 0.174 \\
Rényi--Rényi     & 0.129 & 0.130 & 0.129 & 0.482 & 0.129 \\
TV--TV           & 0.200 & 0.200 & 0.200 & 0.200 & 0.200 \\
\bottomrule
\end{tabular}
\caption{Representative Player~1 action distributions under different 
regularizer pairs $(f_1,f_2)$ in the rank-$2$ game. 
Mixed regularizers (e.g., Entropy--TV, L2--Entropy, L2--TV) visibly 
shift mass allocation compared to homogeneous choices.}
\label{tab:rankk-distributions}
\end{table*}

\begin{figure}[h!]
  \centering
  \begin{subfigure}[t]{0.48\linewidth}
    \centering
    \includegraphics[width=\linewidth]{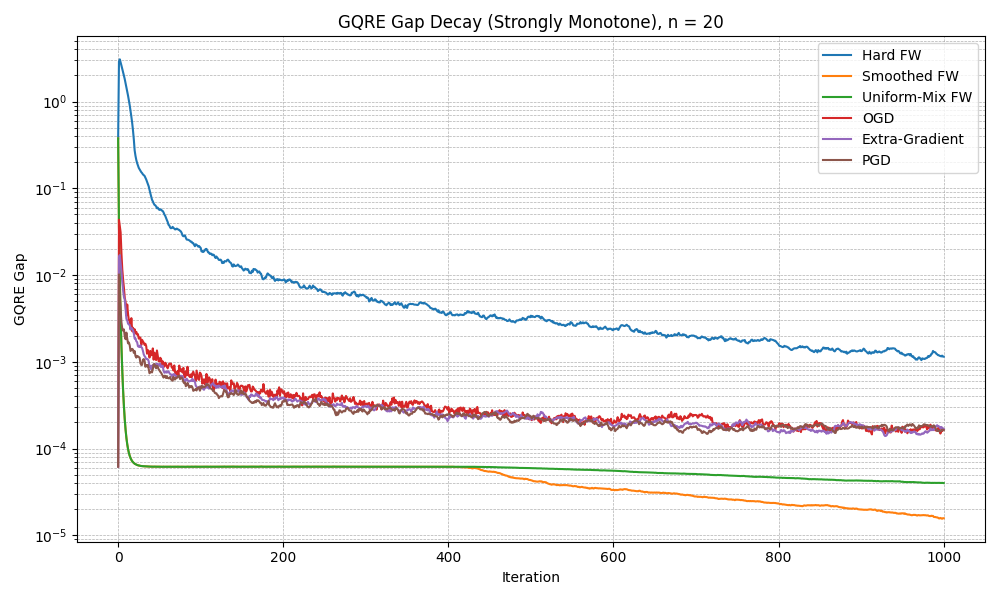}
    \caption{Strongly monotone game ($n=20$).}
  \end{subfigure}
  \hfill
  \begin{subfigure}[t]{0.48\linewidth}
    \centering
    \includegraphics[width=\linewidth]{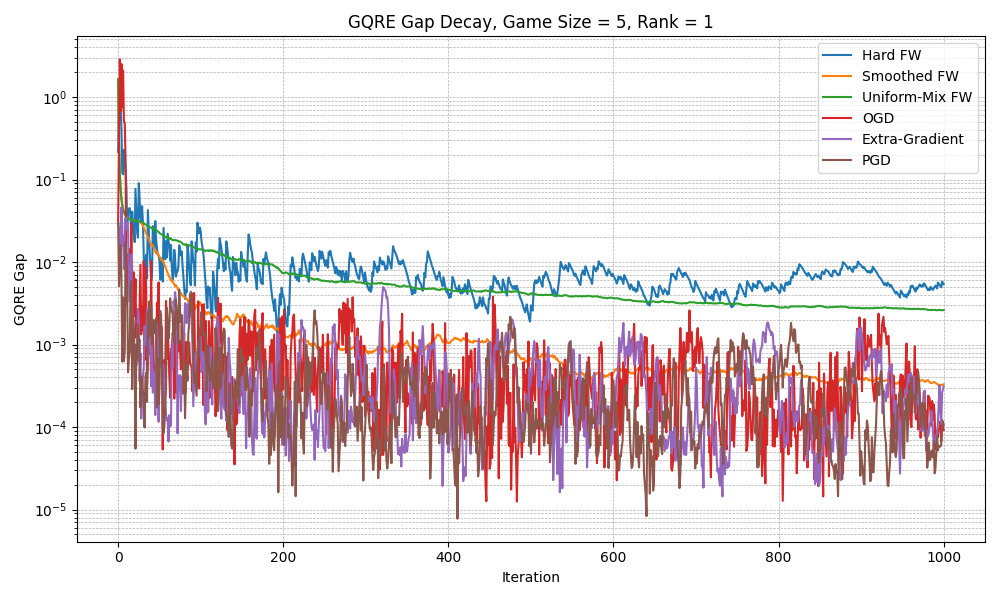}
    \caption{Rank-1 game of size $5$.}
  \end{subfigure}

  \caption{GQRE gap decay in two classes of games. 
  }
  \label{fig:gqre_gap_allschemes}
\end{figure}

In Figure~\ref{fig:gqre_gap_allschemes} we compare the five schemes in terms of the GQRE gap decay for two games from the game families we described above---the first, a strongly monotone game of size 20, and the second, a rank-1 game of size 5. Not only does Algorithm~\ref{algo:fw-decentralized} perform better than existing schemes, but due to the projection step it also outperforms the Uniform-Mix Smoothened Frank-Wolfe algorithm.

\begin{figure}[h]
    \centering
  \begin{subfigure}[t]{0.48\linewidth}
    \centering
        \includegraphics[width=\linewidth]{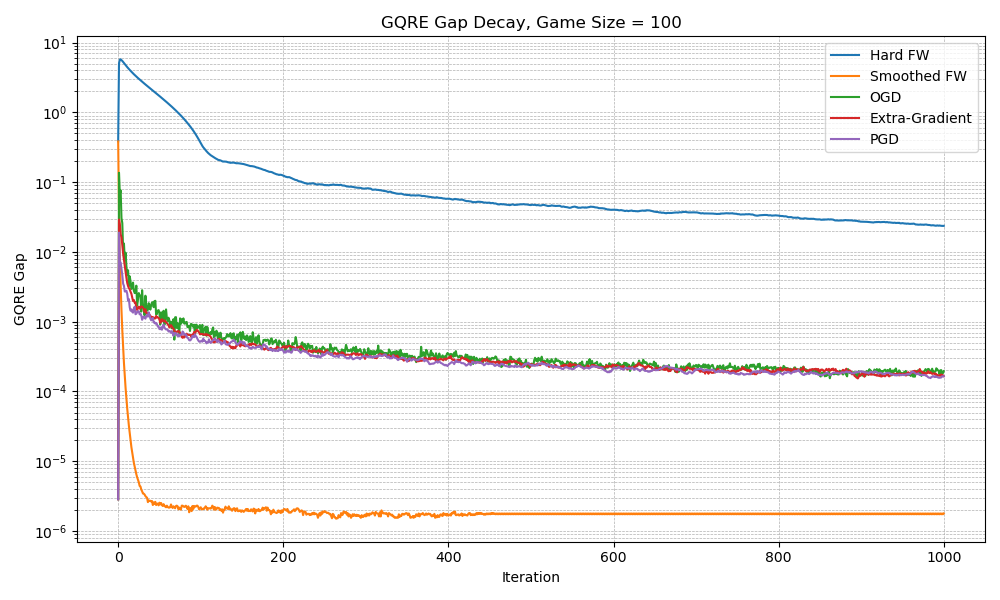}
    \caption{Strongly monotone game ($n=100$).}
   \end{subfigure} 
   \hfill
     \begin{subfigure}[t]{0.48\linewidth}
     \centering
     \includegraphics[width=\linewidth]{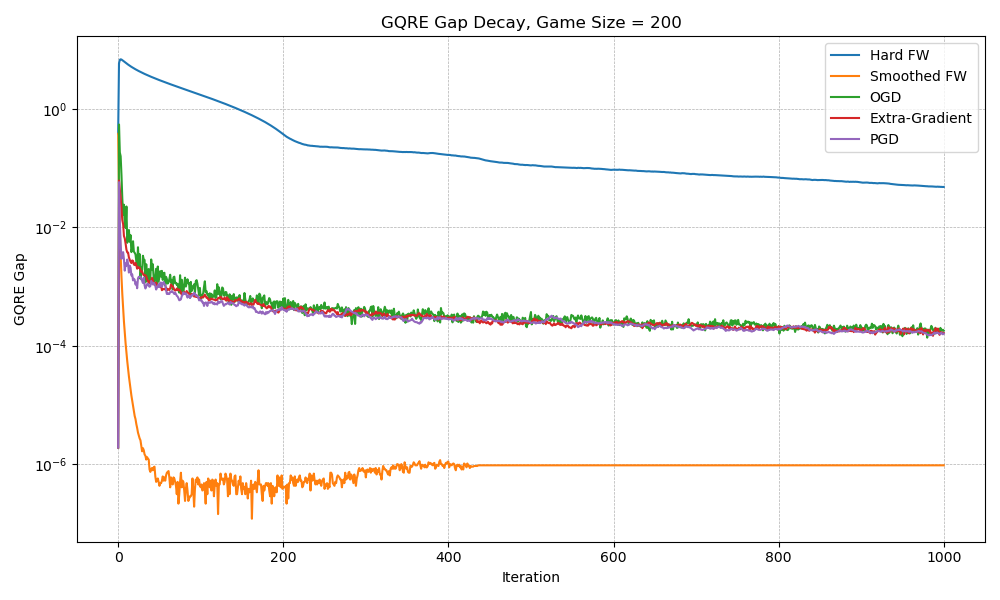}
    \caption{Strongly monotone game ($n=200$).}
    \end{subfigure}

    \centering
        \caption{GQRE gap decay for strongly monotone games with sizes $n = 100, 200$.}
        \label{fig:monotone_lgames}
        
 \end{figure}

\begin{figure}[h]
        \centering
   \begin{subfigure}[t]{0.48\linewidth}
    \centering
    \includegraphics[width=\linewidth]{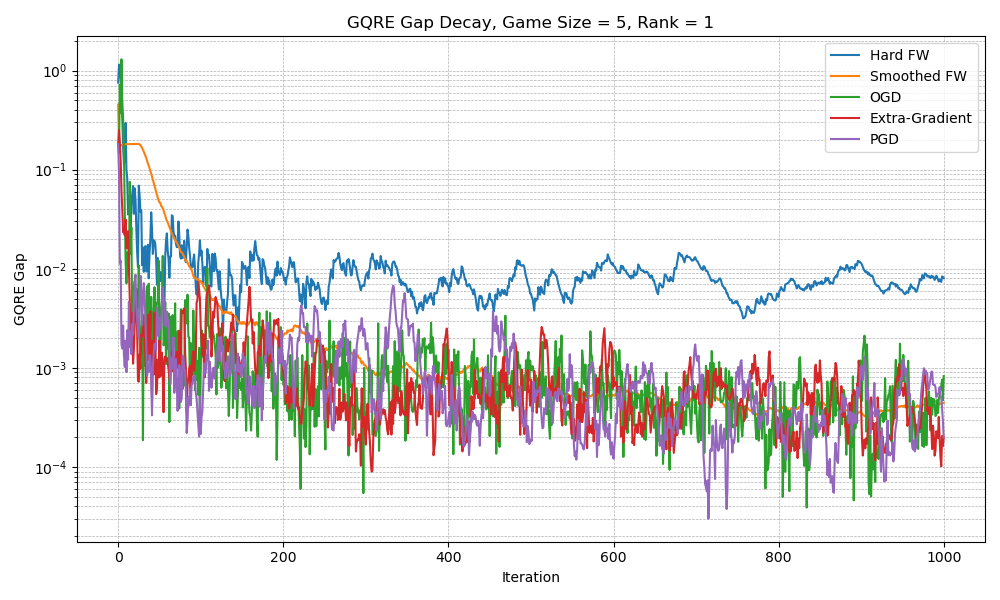}
    \caption{Rank-1 game.}
    \end{subfigure}
  \hfill
    \begin{subfigure}[t]{0.48\linewidth}
    \centering
    \includegraphics[width=\linewidth]{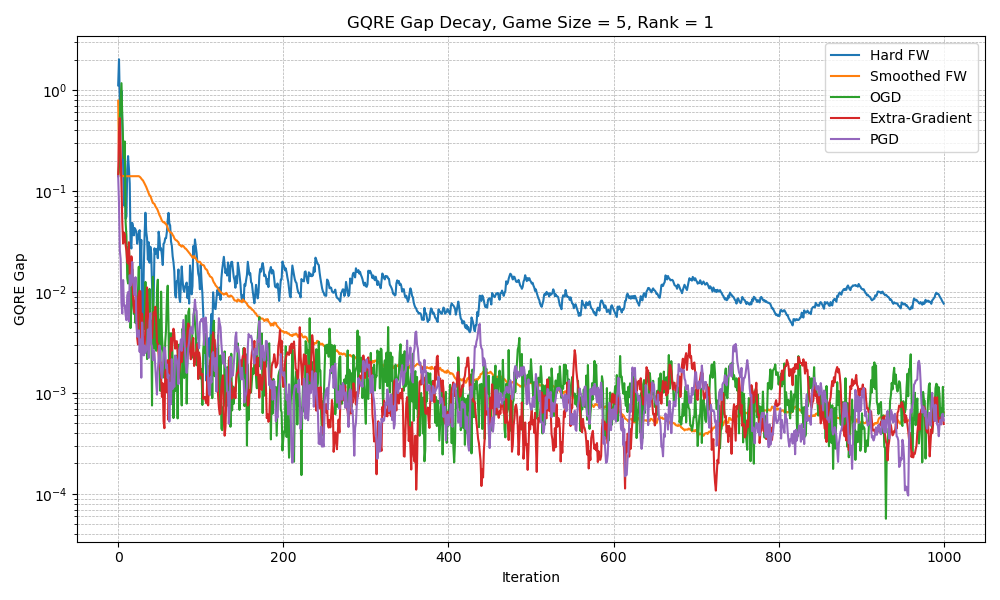}
    \caption{Rank-3 game.}
   \end{subfigure}

    \caption{GQRE gap decay for rank-$k$ general-sum games with size $m = 5$ and ranks $k = 1, 3$.}
    \label{fig:rankx_games}
\end{figure}

Figure \ref{fig:monotone_lgames} shows GQRE gap for the strongly monotone games with game sizes of 100 and 200, Again, we see that Algorithm \ref{algo:fw-decentralized} outperforms all the other algorithms by a wide margin. Figure \ref{fig:rankx_games} shows GQRE gap for rank-$k$ general-sum games for $k=1,3$. Here, we see that our algorithm performs similar to the other baselines, but is much more stable. 

To understand the potential benefit of assigning 
different $f_i$ functions to different agents, we ran simulations with heterogeneous 
regularizers with $\lambda=1$ and Algorithm~\ref{algo:fw-decentralized}. Specifically, for the rank-$k$ game of size $5$, we considered all 
pairwise combinations of four divergences: Shannon entropy (KL), Rényi ($\alpha=0.5$), 
$0.5\| \cdot \|_2^2$ relative to uniform, and the total variation (TV) distance to 
uniform. 

Table~\ref{tab:rankk-distributions} reports representative equilibrium distributions 
for Player~1 under each pair $(f_1,f_2)$. Note that the choice of regularizer shapes the action probabilities: entropy and L2 spread mass more evenly across actions, Rényi accentuates a single action, and TV concentrates very close to uniform. More results are presented in Appendix~\ref{app:add_num_sim}.

In Appendix~\ref{app:add_num_sim} we present numerical results for more games---matching pennies, other strongly monotone games, and rank-$k$ games (ranks $\le 4$). We also study Jordan's three-player two-action game from \citep{jordan1993three}, which has a unique NE, but which cannot be computed using just gradients~\citep{hart2003uncoupled}; this issue shows up for Algorithm~\ref{algo:fw-decentralized} via stability when Assumption~\ref{assmpt:diagonal-dominance} does not hold. Finally, impact of heterogeneous choice of $f_i$\/s by agents is also studied.

Thus, the overall conclusion from this set of results is that the Smoothed FW algorithm we propose is quite effective at computing the GQRE equilibrium in decentralized setting wherein we have access to bandit-feedback of the gradients via a simulator oracle, and each agent adapts independently. 

\section{Conclusions}

In this paper, we study the generalized quantal response equilibrium (GQRE), to predict outcomes of games with bounded-rational players learning independently, and provide sufficient conditions for existence of GQRE. We then introduced a decentralized Smoothened Frank--Wolfe algorithm---Algorithm~\ref{algo:fw-decentralized}---that can be used to compute the GQRE. Through an extensive set of numerical experiments, we showed that it is remarkably good at finding it. 

\BibliographyProjectFile{shared/ref}{../shared/ref}

\onecolumn 
\appendix


\section{Proof of Theorem~\ref{thm:non-asymptotic-fw}}\label{app:proof}

We reiterate the statement of Theorem~\ref{thm:non-asymptotic-fw} below.
\begin{theorem}[No-regret guarantee]
\label{thm:non-asymptotic-fw-new}
Under Assumptions~\ref{assmpt:lipschitz-continuity} and~\ref{assmpt:diagonal-dominance}, and at time-step $t$ step-size $\gamma_t=\tfrac{1}{t+1}$, $\epsilon_t=\tfrac{1}{(t+1)\max_{i}|A_i|}$ and the number of samples $M$ from the simulator chosen as $M_t=\lceil \tfrac{1}{\epsilon_t \gamma_t^2}\rceil$, the expected gap value of Algorithm~\ref{algo:fw-decentralized} at time $T\geq 1$ satisfies: 
\[
R(T) \le \frac{C_7}{T} + \frac{C_{6}\log(T)}{T},
\]
and further, $\lim_{T\rightarrow\infty} V(\boldsymbol{\pi}(T))=0$ in probability.
\end{theorem}

Reminder that 
\begin{align*}
V_i(\boldsymbol{\pi}) & =\eta \log\left(\langle \pi_i, \exp(F_i(\boldsymbol{\pi})/\eta) \rangle \right) - \langle \pi_i, F_i(\boldsymbol{\pi}) \rangle, \text{and } 
V(\boldsymbol{\pi})  = \sum_{i\in[N]} V_i(\boldsymbol{\pi}).
\end{align*}

We have for all $i\in[N]$, 
\begin{align*}
\nabla_{\pi_i} V_i(\boldsymbol{\pi}) & = \eta \frac{\exp(F_i(\boldsymbol{\pi})/\eta)}{\langle \pi_i, \exp(F_i(\boldsymbol{\pi})/\eta) \rangle} - F_i(\boldsymbol{\pi}) \\
& \qquad + \nabla_{\pi_i} F_i(\boldsymbol{\pi})^T\left( s_i^*(\boldsymbol{\pi},F_i(\boldsymbol{\pi})) - \pi_i \right)  \\
\forall j\in -i,\ \nabla_{\pi_j} V_i(\boldsymbol{\pi}) & = \nabla_{\pi_j} F_i(\boldsymbol{\pi})^T\left( s_i^*(\boldsymbol{\pi},F_i(\boldsymbol{\pi})) - \pi_i \right) .
\end{align*}

Our ``RL" simulator provides noisy gradient $\boldsymbol{\delta}(\boldsymbol{\pi})+F(\boldsymbol{\pi})$ (with $\boldsymbol{\delta}(\boldsymbol{\pi})$ being zero mean), which we use in the algorithm 
instead of $F(\boldsymbol{\pi})$. This gives $\boldsymbol{s}^*(\boldsymbol{\pi}, F(\boldsymbol{\pi})+\boldsymbol{\delta}(\boldsymbol{\pi}))$ instead of $\boldsymbol{s}^*(\boldsymbol{\pi}, F(\boldsymbol{\pi}))$ in the noisy case. Further, the projection brings it to $\boldsymbol{s}_{\epsilon}^*(\boldsymbol{\pi}, F(\boldsymbol{\pi})+\boldsymbol{\delta}(\boldsymbol{\pi}))$.

Based on Assumption~\ref{assmpt:lipschitz-continuity}, the bound we will use is 
\begin{align*}
V(\boldsymbol{\pi}(t+1)) - V(\boldsymbol{\pi}(t) ) & \leq \langle \nabla_{\boldsymbol{\pi}} V(\boldsymbol{\pi}(t)), \boldsymbol{\pi}(t+1)) - \boldsymbol{\pi}(t))\rangle + \frac{C}{2} \|\boldsymbol{\pi}(t+1)) - \boldsymbol{\pi}(t) \|_2^2 \numberthis \label{eq:ineqVgen}\\
& = \gamma_t \langle \nabla_{\boldsymbol{\pi}} V(\boldsymbol{\pi}(t)), \boldsymbol{s}_{\epsilon}^*(\boldsymbol{\pi}(t), F(\boldsymbol{\pi}(t))+\boldsymbol{\delta}(\boldsymbol{\pi}(t)))-\boldsymbol{\pi}(t)\rangle \\
& \quad + \frac{C \gamma_t^2}{2} \|\boldsymbol{s}_{\epsilon}^*(\boldsymbol{\pi}(t), F(\boldsymbol{\pi}(t)+\boldsymbol{\delta}(\boldsymbol{\pi}(t))))-\boldsymbol{\pi}(t) \|_2^2, \numberthis \label{eq:ineqV}
\end{align*}
where the second representation uses our update rule. We prove this bound later on in Lemma~\ref{lem:Csmooth}.

We will focus on the first term on the RHS of \eqref{eq:ineqV} after dividing by the $\gamma_t$ term, namely,
\begin{align*}
& \langle \nabla_{\boldsymbol{\pi}} V(\boldsymbol{\pi}(t)), \boldsymbol{s}_{\epsilon}^*(\boldsymbol{\pi}(t), F(\boldsymbol{\pi}(t))+\boldsymbol{\delta}(\boldsymbol{\pi}(t)))-\boldsymbol{\pi}(t)\rangle \\
& = \sum_{i\in[N]} \langle \nabla_{\boldsymbol{\pi}} V_i(\boldsymbol{\pi}(t)), \boldsymbol{s}_{\epsilon}^*(\boldsymbol{\pi}(t), F(\boldsymbol{\pi}(t))+\boldsymbol{\delta}(\boldsymbol{\pi}(t)))-\boldsymbol{\pi}(t)\rangle \\
& = \sum_{i\in[N]} \langle \nabla_{\boldsymbol{\pi}} V_i(\boldsymbol{\pi}(t)), \boldsymbol{s}_{\epsilon}^*(\boldsymbol{\pi}(t), F(\boldsymbol{\pi}(t))+\boldsymbol{\delta}(\boldsymbol{\pi}(t)))-\boldsymbol{s}^*(\boldsymbol{\pi}(t), F(\boldsymbol{\pi}(t))+\boldsymbol{\delta}(\boldsymbol{\pi}(t)))\rangle \\
& \qquad\qquad +  \langle \nabla_{\boldsymbol{\pi}} V_i(\boldsymbol{\pi}(t)), \boldsymbol{s}^*(\boldsymbol{\pi}(t), F(\boldsymbol{\pi}(t))+\boldsymbol{\delta}(\boldsymbol{\pi}(t)))-\boldsymbol{\pi}(t)\rangle.
\end{align*}
The first term is small here as it is dependent on $\boldsymbol{s}_{\epsilon}^*(\boldsymbol{\pi}(t), F(\boldsymbol{\pi}(t))+\boldsymbol{\delta}(\boldsymbol{\pi}(t)))-\boldsymbol{s}^*(\boldsymbol{\pi}(t), F(\boldsymbol{\pi}(t))+\boldsymbol{\delta}(\boldsymbol{\pi}(t)))$, and this is a function of $\epsilon_t$ via our projection; this will also holds for the schemes of \citep{kocak2014efficient,heliou2017learning}. We will include it in our statement of the result in the end, so we will focus on the second term, namely,
\begin{align*}
& \sum_{i\in[N]} \langle \nabla_{\boldsymbol{\pi}} V_i(\boldsymbol{\pi}(t)), \boldsymbol{s}^*(\boldsymbol{\pi}(t), F(\boldsymbol{\pi}(t))+\boldsymbol{\delta}(\boldsymbol{\pi}(t)))-\boldsymbol{\pi}(t)\rangle \\
& = \sum_{i\in[N]} \left\langle  \eta \frac{\exp(F_i(\boldsymbol{\pi}(t))/\eta)}{\langle \pi_i(t), \exp(F_i(\boldsymbol{\pi}(t))/\eta) \rangle} - F_i(\boldsymbol{\pi}(t)), s_i^*(\boldsymbol{\pi}(t), F(\boldsymbol{\pi}(t))+\boldsymbol{\delta}(\boldsymbol{\pi}(t)))-\pi_i(t)\right\rangle \\
& \qquad + ( \boldsymbol{s}^*(\boldsymbol{\pi}(t), F(\boldsymbol{\pi}(t)))-\boldsymbol{\pi}(t) )^T H(\boldsymbol{\pi}(t))  (\boldsymbol{s}^*(\boldsymbol{\pi}(t), F(\boldsymbol{\pi}(t))+\boldsymbol{\delta}(\boldsymbol{\pi}(t)))-\boldsymbol{\pi}(t))\\
& = \sum_{i\in[N]} \left\langle  \eta \frac{\exp(F_i(\boldsymbol{\pi}(t))/\eta)}{\langle \pi_i(t), \exp(F_i(\boldsymbol{\pi}(t))/\eta) \rangle} - F_i(\boldsymbol{\pi}(t)), s_i^*(\boldsymbol{\pi}(t), F(\boldsymbol{\pi}(t))+\boldsymbol{\delta}(\boldsymbol{\pi}(t)))-s_i^*(\boldsymbol{\pi}(t), F(\boldsymbol{\pi}(t)))\right\rangle \\
& \qquad + ( \boldsymbol{s}^*(\boldsymbol{\pi}(t), F(\boldsymbol{\pi}(t)))-\boldsymbol{\pi}(t) )^T H(\boldsymbol{\pi}(t))  (\boldsymbol{s}^*(\boldsymbol{\pi}(t), F(\boldsymbol{\pi}(t))+\boldsymbol{\delta}(\boldsymbol{\pi}(t)))-\boldsymbol{\pi}(t))\\
&\qquad  + \sum_{i\in[N]} \left\langle  \eta \frac{\exp(F_i(\boldsymbol{\pi}(t))/\eta)}{\langle \pi_i(t), \exp(F_i(\boldsymbol{\pi})/\eta) \rangle} - F_i(\boldsymbol{\pi}), s_i^*(\boldsymbol{\pi}(t), F(\boldsymbol{\pi}(t)))-\pi_i(t)\right\rangle \numberthis \label{eq:thirdterm} 
\end{align*}
By Legendre-Fenchel duality, we have
\begin{align*}
V_i(\boldsymbol{\pi}) + \langle \pi_i , F_i(\boldsymbol{\pi})\rangle & = \max_{s\in \Delta(A_i)} \langle s, F_i(\boldsymbol{\pi}) \rangle - \eta \mathrm{KL}( s | \pi_i) \\
& = \langle s_i^*(\boldsymbol{\pi}, F(\boldsymbol{\pi}), F_i(\boldsymbol{\pi}) \rangle - \eta \mathrm{KL}( s_i^*(\boldsymbol{\pi}, F(\boldsymbol{\pi}) | \pi_i).
\end{align*}
Thus, the third term in \eqref{eq:thirdterm} yields
\begin{align*}
& \sum_{i\in[N]} \left\langle  \eta \frac{\exp(F_i(\boldsymbol{\pi}(t))/\eta)}{\langle \pi_i(t), \exp(F_i(\boldsymbol{\pi}(t))/\eta) \rangle} - F_i(\boldsymbol{\pi}(t)), s_i^*(\boldsymbol{\pi}(t), F(\boldsymbol{\pi}(t)))-\pi_i(t)\right\rangle \\
& = \sum_{i\in[N]} \left\langle  \eta \frac{\exp(F_i(\boldsymbol{\pi}(t))/\eta)}{\langle \pi_i(t), \exp(F_i(\boldsymbol{\pi}(t))/\eta) \rangle} , s_i^*(\boldsymbol{\pi}(t), F(\boldsymbol{\pi}(t)))-\pi_i(t)\right\rangle \\
& \quad - \sum_{i\in[N]} \left\langle   F_i(\boldsymbol{\pi}), s_i^*(\boldsymbol{\pi}(t), F(\boldsymbol{\pi}(t)))-\pi_i(t)\right\rangle \\
& = \sum_{i\in[N]}  \left\langle  \eta \frac{\exp(F_i(\boldsymbol{\pi}(t))/\eta)}{\langle \pi_i(t), \exp(F_i(\boldsymbol{\pi}(t))/\eta) \rangle} , s_i^*(\boldsymbol{\pi}(t), F(\boldsymbol{\pi}(t)))-\pi_i(t)\right\rangle \\
& \quad - \sum_{i\in[N]} \left( V_i(\boldsymbol{\pi}(t)) + \eta \mathrm{KL}( s_i^*(\boldsymbol{\pi}(t), F(\boldsymbol{\pi}(t))) | \pi_i(t)) \right)\\
& = - V(\boldsymbol{\pi}(t)) + \sum_{i\in[N]}  \left\langle  \eta \frac{\exp(F_i(\boldsymbol{\pi}(t))/\eta)}{\langle \pi_i(t), \exp(F_i(\boldsymbol{\pi}(t))/\eta) \rangle} , s_i^*(\boldsymbol{\pi}(t), F(\boldsymbol{\pi}(t)))-\pi_i(t)\right\rangle \\
& \quad - \eta \sum_{i\in[N]} \mathrm{KL}( s_i^*(\boldsymbol{\pi}(t), F(\boldsymbol{\pi}(t))) | \pi_i(t))
\end{align*}
Inserting the revised form of the third term into \eqref{eq:thirdterm} yields
\begin{align}
& \sum_{i\in[N]} \langle \nabla_{\boldsymbol{\pi}} V_i(\boldsymbol{\pi}(t)), \boldsymbol{s}^*(\boldsymbol{\pi}(t), F(\boldsymbol{\pi}(t))+\boldsymbol{\delta}(\boldsymbol{\pi}(t)))-\boldsymbol{\pi}(t)\rangle \notag \\ 
& = - V(\boldsymbol{\pi}(t)) - \sum_{i\in[N]} \left\langle  F_i(\boldsymbol{\pi}(t)), s_i^*(\boldsymbol{\pi}(t), F(\boldsymbol{\pi}(t))+\boldsymbol{\delta}(\boldsymbol{\pi}(t)))-s_i^*(\boldsymbol{\pi}(t), F(\boldsymbol{\pi}(t)))\right\rangle \notag \\
&  \quad + ( \boldsymbol{s}^*(\boldsymbol{\pi}(t), F(\boldsymbol{\pi}(t)))-\boldsymbol{\pi}(t) )^T H(\boldsymbol{\pi}(t))  (\boldsymbol{s}^*(\boldsymbol{\pi}(t), F(\boldsymbol{\pi}(t))+\boldsymbol{\delta}(\boldsymbol{\pi}(t)))-\boldsymbol{\pi}(t)) \notag\\
& \quad + \eta\sum_{i\in[N]} \left\langle   \frac{\exp(F_i(\boldsymbol{\pi})/\eta)}{\langle \pi_i(t), \exp(F_i(\boldsymbol{\pi})/\eta) \rangle} , s_i^*(\boldsymbol{\pi}(t), F(\boldsymbol{\pi}(t))+\boldsymbol{\delta}(\boldsymbol{\pi}(t)))-\pi_i(t)\right\rangle \notag\\
& \quad - \eta \sum_{i\in[N]} \mathrm{KL}( s_i^*(\boldsymbol{\pi}(t), F(\boldsymbol{\pi}(t))) | \pi_i(t)) \label{eq:eq1}\\
& \leq  - V(\boldsymbol{\pi}(t)) - \sum_{i\in[N]} \left\langle  F_i(\boldsymbol{\pi}(t)), s_i^*(\boldsymbol{\pi}(t), F(\boldsymbol{\pi}(t))+\boldsymbol{\delta}(\boldsymbol{\pi}(t)))-s_i^*(\boldsymbol{\pi}(t), F(\boldsymbol{\pi}(t)))\right\rangle \notag \\
& \quad + ( \boldsymbol{s}^*(\boldsymbol{\pi}(t), F(\boldsymbol{\pi}(t)))-\boldsymbol{\pi}(t) )^T H(\boldsymbol{\pi}(t))  (\boldsymbol{s}^*(\boldsymbol{\pi}(t), F(\boldsymbol{\pi}(t))+\boldsymbol{\delta}(\boldsymbol{\pi}(t)))-\boldsymbol{\pi}(t))\notag \\
& \quad + \eta \sum_{i\in[N]} \Big( \mathrm{KL}(\boldsymbol{s}^*(\boldsymbol{\pi}(t), F(\boldsymbol{\pi}(t))) | \boldsymbol{s}^*(\boldsymbol{\pi}(t), F(\boldsymbol{\pi}(t))+\boldsymbol{\delta}(\boldsymbol{\pi}(t))) ) \notag\\
& \qquad \qquad \qquad\qquad\qquad -  2\mathrm{KL}( s_i^*(\boldsymbol{\pi}(t), F(\boldsymbol{\pi}(t))) | \pi_i(t)) \Big) \label{eq:ineq1}
\end{align}
\begin{remark}
From the bound in \eqref{eq:ineq1} it is clear that the exact (or noiseless) gradient setting has negative drift, since $F(\boldsymbol{\pi}(t))+\boldsymbol{\delta}(\boldsymbol{\pi}(t))=F(\boldsymbol{\pi}(t))$ in this setting. Hence, this algorithm is a good computational option when the payoff matrix is known. Assumption~\ref{assmpt:diagonal-dominance} is used to show that the quadratic forms are negative.
\end{remark}

The inequality in \eqref{eq:ineq1} is obtained the using the fact $\mathrm{KL}(p|q)$ for two distributions $p$ and $q$ is convex in $q$, and hence, satisfies the gradient inequality: for three distributions $p$, $q$ and $q_1$, we have
\begin{align*}
\mathrm{KL}(p|q_1) & \geq \mathrm{KL}(p|q) +   \nabla_q \mathrm{KL}(p|q)^T (q_1-q), \text{and} 
 \left(\nabla_q \mathrm{KL}(p|q)\right)_k  = \frac{p_k}{q_k},\ \forall k\in [K],
\end{align*}
where we assume that the distributions are in a $K$-valued discrete space (and, sufficient for our purposes, both $q$  and $q_1$ have full support). Then, we set the following:
\begin{align*}
p&=\boldsymbol{s}^*(\boldsymbol{\pi}(t), F(\boldsymbol{\pi}(t))),\ q=\pi_i(t), \text{ and } q_1=\boldsymbol{s}^*(\boldsymbol{\pi}(t), F(\boldsymbol{\pi}_{e}(t))).
\end{align*}
Using the expression for $\boldsymbol{s}^*(\boldsymbol{\pi}(t), F(\boldsymbol{\pi}(t)))$, this yields
\begin{align*}
& \mathrm{KL}(s_i^*(\boldsymbol{\pi}(t), F(\boldsymbol{\pi}(t))) | s_i^*(\boldsymbol{\pi}(t), F(\boldsymbol{\pi}(t))+\boldsymbol{\delta}(\boldsymbol{\pi}(t))) )\\
&  \geq \mathrm{KL}(s_i^*(\boldsymbol{\pi}(t), F(\boldsymbol{\pi}(t))) | \pi_i(t) ) + \left\langle   \frac{\exp(F_i(\boldsymbol{\pi})/\eta)}{\langle \pi_i(t), \exp(F_i(\boldsymbol{\pi})/\eta) \rangle} , s_i^*(\boldsymbol{\pi}(t), F(\boldsymbol{\pi}(t))+\boldsymbol{\delta}(\boldsymbol{\pi}(t)))-\pi_i(t)\right\rangle
\end{align*}

Consider \eqref{eq:eq1} where the RHS is given by
\begin{align*}
\mathrm{RHS} & = - V(\boldsymbol{\pi}(t)) - \left\langle  F(\boldsymbol{\pi}(t)), \boldsymbol{s}^*(\boldsymbol{\pi}(t), F(\boldsymbol{\pi}(t))+\boldsymbol{\delta}(\boldsymbol{\pi}(t)))-\boldsymbol{s}^*(\boldsymbol{\pi}(t), F(\boldsymbol{\pi}(t)))\right\rangle \notag \\
&  \quad + ( \boldsymbol{s}^*(\boldsymbol{\pi}(t), F(\boldsymbol{\pi}(t)))-\boldsymbol{\pi}(t) )^T H(\boldsymbol{\pi}(t))  (\boldsymbol{s}^*(\boldsymbol{\pi}(t), F(\boldsymbol{\pi}(t))+\boldsymbol{\delta}(\boldsymbol{\pi}(t)))-\boldsymbol{\pi}(t)) \notag\\
& \quad + \eta\sum_{i\in[N]} \left\langle   \frac{\exp(F_i(\boldsymbol{\pi}(t))/\eta)}{\langle \pi_i(t), \exp(F_i(\boldsymbol{\pi}(t))/\eta) \rangle} , s_i^*(\boldsymbol{\pi}(t), F(\boldsymbol{\pi}(t))+\boldsymbol{\delta}(\boldsymbol{\pi}(t)))-\pi_i(t)\right\rangle \notag \\
& \quad - \eta \sum_{i\in[N]} \mathrm{KL}( s_i^*(\boldsymbol{\pi}(t), F(\boldsymbol{\pi}(t))) | \pi_i(t))\\
& = - V(\boldsymbol{\pi}(t)) + ( \boldsymbol{s}^*(\boldsymbol{\pi}(t), F(\boldsymbol{\pi}(t)))-\boldsymbol{\pi}(t) )^T H(\boldsymbol{\pi}(t))  (\boldsymbol{s}^*(\boldsymbol{\pi}(t), F(\boldsymbol{\pi}(t)))-\boldsymbol{\pi}(t)) \\
& \quad + \eta\sum_{i\in[N]} \left\langle   \frac{\exp(F_i(\boldsymbol{\pi}(t))/\eta)}{\langle \pi_i(t), \exp(F_i(\boldsymbol{\pi}(t))/\eta) \rangle} , s_i^*(\boldsymbol{\pi}(t), F(\boldsymbol{\pi}(t)))-\pi_i(t)\right\rangle \\
& \quad - \eta \sum_{i\in[N]} \mathrm{KL}( s_i^*(\boldsymbol{\pi}(t), F(\boldsymbol{\pi}(t))) | \pi_i(t))\\
& \quad + ( \boldsymbol{s}^*(\boldsymbol{\pi}(t), F(\boldsymbol{\pi}(t)))-\boldsymbol{\pi}(t) )^T H(\boldsymbol{\pi}(t))  (\boldsymbol{s}^*(\boldsymbol{\pi}(t), F(\boldsymbol{\pi}(t))+\boldsymbol{\delta}(\boldsymbol{\pi}(t)))-\boldsymbol{s}^*(\boldsymbol{\pi}(t), F(\boldsymbol{\pi}(t))))\\
& \quad - \left\langle  F(\boldsymbol{\pi}(t)), \boldsymbol{s}^*(\boldsymbol{\pi}(t), F(\boldsymbol{\pi}(t))+\boldsymbol{\delta}(\boldsymbol{\pi}(t)))-\boldsymbol{s}^*(\boldsymbol{\pi}(t), F(\boldsymbol{\pi}(t)))\right\rangle \\
& \quad + \eta\sum_{i\in[N]} \left\langle   \frac{\exp(F_i(\boldsymbol{\pi})/\eta)}{\langle \pi_i(t), \exp(F_i(\boldsymbol{\pi})/\eta) \rangle} , s_i^*(\boldsymbol{\pi}(t), F(\boldsymbol{\pi}(t))+\boldsymbol{\delta}(\boldsymbol{\pi}(t)))-s_i^*(\boldsymbol{\pi}(t), F(\boldsymbol{\pi}(t)))\right\rangle \\
&=  G(\boldsymbol{\pi}(t)) + \langle L(\boldsymbol{\pi}(t)), \boldsymbol{s}^*(\boldsymbol{\pi}(t), F(\boldsymbol{\pi}(t))+\boldsymbol{\delta}(\boldsymbol{\pi}(t)))-\boldsymbol{s}^*(\boldsymbol{\pi}(t), F(\boldsymbol{\pi}(t)))\rangle
\end{align*}
where
\begin{align*}
 G(\boldsymbol{\pi}(t)) & = - V(\boldsymbol{\pi}(t)) + ( \boldsymbol{s}^*(\boldsymbol{\pi}(t), F(\boldsymbol{\pi}(t)))-\boldsymbol{\pi}(t) )^T H(\boldsymbol{\pi}(t))  (\boldsymbol{s}^*(\boldsymbol{\pi}(t), F(\boldsymbol{\pi}(t)))-\boldsymbol{\pi}(t)) \\
& \quad + \eta\sum_{i\in[N]} \left\langle   \frac{\exp(F_i(\boldsymbol{\pi}(t))/\eta)}{\langle \pi_i(t), \exp(F_i(\boldsymbol{\pi}(t))/\eta) \rangle} , s_i^*(\boldsymbol{\pi}(t), F(\boldsymbol{\pi}(t)))-\pi_i(t)\right\rangle \\
& \quad - \eta \sum_{i\in[N]} \mathrm{KL}( s_i^*(\boldsymbol{\pi}(t), F(\boldsymbol{\pi}(t))) | \pi_i(t))\\
& \leq - V(\boldsymbol{\pi}(t)),
\end{align*}
but the other negative terms will be useful to drive other quantities negative---some of them arise from Assumption~\ref{assmpt:diagonal-dominance} regarding the quadratic forms. Defining vector $\boldsymbol{\lambda}(\boldsymbol{\pi})$ by setting the $i^{\mathrm{th}}$ block to be
\begin{align*}
\lambda_i(\boldsymbol{\pi}) = \eta \frac{\exp(F_i(\boldsymbol{\pi})/\eta)}{\langle \pi_i, \exp(F_i(\boldsymbol{\pi})/\eta) \rangle},
\end{align*}
we have 
\begin{align*}
L(\boldsymbol{\pi}(t)) = -F(\boldsymbol{\pi}(t)) + H^T(\boldsymbol{\pi}(t)) ( \boldsymbol{s}^*(\boldsymbol{\pi}(t), F(\boldsymbol{\pi}(t)))-\boldsymbol{\pi}(t) )  + \boldsymbol{\lambda}(\boldsymbol{\pi}(t))
\end{align*}

Now collecting everything we can replace \eqref{eq:ineqV} by
\begin{align*}
& V(\boldsymbol{\pi}(t+1)) -V(\boldsymbol{\pi}(t)) \\
& \leq \gamma_t   G(\boldsymbol{\pi}(t)) +  \gamma_t \langle L(\boldsymbol{\pi}(t)), \boldsymbol{s}^*(\boldsymbol{\pi}(t), F(\boldsymbol{\pi}(t))+\boldsymbol{\delta}(\boldsymbol{\pi}(t)))-\boldsymbol{s}^*(\boldsymbol{\pi}(t), F(\boldsymbol{\pi}(t)))\rangle
\\ & \quad + \gamma_t \sum_{i\in[N]} \langle \nabla_{\boldsymbol{\pi}} V_i(\boldsymbol{\pi}(t)), s_{i,\epsilon}^*(\boldsymbol{\pi}(t), F_i(\boldsymbol{\pi}(t))+\boldsymbol{\delta}(\boldsymbol{\pi}(t)))-s_i^*(\boldsymbol{\pi}(t), F_i(\boldsymbol{\pi}(t))+\boldsymbol{\delta}(\boldsymbol{\pi}(t)))\rangle \\
& \quad + \frac{C \gamma_t^2}{2} \|\boldsymbol{s}_{\epsilon}^*(\boldsymbol{\pi}(t), F(\boldsymbol{\pi}(t)+\boldsymbol{\delta}(\boldsymbol{\pi}(t))))-\boldsymbol{\pi}(t) \|_2^2\\
& \leq \gamma_t   G(\boldsymbol{\pi}(t)) +  \gamma_t \langle L(\boldsymbol{\pi}(t)), s_i^*(\boldsymbol{\pi}(t), F(\boldsymbol{\pi}(t))+\boldsymbol{\delta}(\boldsymbol{\pi}(t)))-s_i^*(\boldsymbol{\pi}(t), F(\boldsymbol{\pi}(t)))\rangle \\
& \quad+ 2 \gamma_t \epsilon_t C_1 + C_2 \gamma_t^2
\end{align*}
Hence, the expected drift is given by
\begin{align*}
& \mE[ V(\boldsymbol{\pi}(t+1)) -V(\boldsymbol{\pi}(t)) | \boldsymbol{\pi}(t) ] \\
& \leq \gamma_t   G(\boldsymbol{\pi}(t)) +  \gamma_t \mE[\langle L(\boldsymbol{\pi}(t)), s_i^*(\boldsymbol{\pi}(t), F(\boldsymbol{\pi}(t))+\boldsymbol{\delta}(\boldsymbol{\pi}(t)))-s_i^*(\boldsymbol{\pi}(t), F(\boldsymbol{\pi}(t)))\rangle | \boldsymbol{\pi}(t)] \\
& \quad+ 2 \gamma_t \epsilon_t C_1 + C_2 \gamma_t^2.
\end{align*}
Applying the Cauchy-Schwartz inequality and then bounding, we get
\begin{align*}
    & \mE[ V(\boldsymbol{\pi}(t+1)) -V(\boldsymbol{\pi}(t)) | \boldsymbol{\pi}(t) ] \\
    & \leq \gamma_t   G(\boldsymbol{\pi}(t)) +  \gamma_t \mE[\langle L(\boldsymbol{\pi}(t)), \boldsymbol{s}^*(\boldsymbol{\pi}(t), F(\boldsymbol{\pi}(t))+\boldsymbol{\delta}(\boldsymbol{\pi}(t)))-\boldsymbol{s}^*(\boldsymbol{\pi}(t), F(\boldsymbol{\pi}(t)))\rangle | \boldsymbol{\pi}(t)] \\
    & \quad + 2 \gamma_t \epsilon_t C_1 + C_2 \gamma_t^2\\
    & = \gamma_t   G(\boldsymbol{\pi}(t)) +  \gamma_t \sum_{i\in[N]}\mE[\langle L_i(\boldsymbol{\pi}(t)), s_i^*(\boldsymbol{\pi}(t), F(\boldsymbol{\pi}(t))+\boldsymbol{\delta}(\boldsymbol{\pi}(t)))-s_i^*(\boldsymbol{\pi}(t), F(\boldsymbol{\pi}(t)))\rangle | \boldsymbol{\pi}(t)] \\
    & \quad + 2 \gamma_t \epsilon_t C_1 + C_2 \gamma_t^2\\
    & \leq  \gamma_t   G(\boldsymbol{\pi}(t)) +  \gamma_t C_2 \eta \sum_{i\in[N]}\sqrt{\mE[\| s_i^*(\boldsymbol{\pi}(t), F_i(\boldsymbol{\pi}(t))+\boldsymbol{\delta}(\boldsymbol{\pi}(t)))-s_i^*(\boldsymbol{\pi}(t), F_i(\boldsymbol{\pi}(t))) \|_2^2]} \\
    & \quad + 2 \gamma_t \epsilon_t C_1 + C_2 \gamma_t^2\\
    & \leq \gamma_t   G(\boldsymbol{\pi}(t)) +  \gamma_t C_2  \sum_{i\in[N]} \sqrt{\mE[\| \delta_i(\boldsymbol{\pi}(t))) \|_2^2]}  + 2 \gamma_t \epsilon_t C_1 + C_2 \gamma_t^2\\
    & \leq \gamma_t   G(\boldsymbol{\pi}(t)) +  \gamma_t C_2  \sum_{i\in[N]} \sqrt{ \sum_{a_i\in A_i} \frac{\lambda_i^2\widetilde{\mathrm{Var}}(a_i,\pi_i,\pi_{-i};u_i)}{M \pi_i(a_i)} } + 2 \gamma_t \epsilon_t C_1 + C_2 \gamma_t^2\\
    & \leq \gamma_t   G(\boldsymbol{\pi}(t)) +  \gamma_t C_3 \frac{1}{\sqrt{M \epsilon_t}}  + 2 \gamma_t \epsilon_t C_1 + C_2 \gamma_t^2,
\end{align*}
where we have used the fact that the softmax function~\cite[Proposition 4]{gao2017properties} is $1/\eta$ Lipschitz in the $l_2$ norm with respect to the input vector---our function $s_i^*(\boldsymbol{\pi}(t), F_i(\boldsymbol{\pi}(t)))$ is a softmax function where the second component is the input vector, and we consider the difference between the values at $F_i(\boldsymbol{\pi}(t))+\delta_i(\boldsymbol{\pi}(t)))$ and $F_i(\boldsymbol{\pi}(t))$. We have also used the variance characterization in \eqref{eqn:gradient_estimate}, and we remind the reader that from \eqref{eq:varterm} we have $\forall a_i \in A_i$,
\begin{align*}
\widetilde{\mathrm{Var}}(a_i,\pi_i,\pi_{-i};u_i) & =\sum_{\boldsymbol{a}_{-i}\in \prod_{j\in -i} A_j} \pi_{-i}(\boldsymbol{a}_{-i}) u_i^2(a_i,\boldsymbol{a}_{-i}) - \pi_i(a_i) u_i^2(a_i,\pi_{-i}) \\
& \leq \sum_{\boldsymbol{a}_{-i}\in \prod_{j\in -i} A_j} \pi_{-i}(\boldsymbol{a}_{-i}) u_i^2(a_i,\boldsymbol{a}_{-i}) \leq 1,
\end{align*}
where we used our assumed upper bound of $1$ on the utilities.

For completeness, we derive the variance characterization in \eqref{eqn:gradient_estimate} by showing the following:
\begin{align*}
    \mE[U_i^2(a_i)] & = \mE\left[ \sum_{m, m^\prime=1}^M u_i(\boldsymbol{A}(m)) u_i(\boldsymbol{A}(m^\prime)) 1_{\{A_i(m)=A_i(m^\prime)=a_i\}}\right]\\
    & = \sum_{m=1}^M \mE\left[ u_i^2(\boldsymbol{A}(m)) 1_{\{A_i(m)=a_i\}}\right] \\
    & \quad +  \sum_{m, m^\prime=1, m\neq m^\prime}^M \mE\left[ u_i(\boldsymbol{A}(m)) u_i(\boldsymbol{A}(m^\prime)) 1_{\{A_i(m)=A_i(m)=a_i\}}\right]\\
    & = M \pi_i(a_i) \sum_{\boldsymbol{a}_{-i}\in \prod_{j\in -i} A_j} \pi_{-i}(\boldsymbol{a}_{-i}) u_i^2(a_i,\boldsymbol{a}_{-i}) + M(M-1) \pi_i^2(a_i) u_i^2(a_i,\pi_{-i}) \\
    & = M \left(\pi_i(a_i)\sum_{\boldsymbol{a}_{-i}\in \prod_{j\in -i} A_j} \pi_{-i}(\boldsymbol{a}_{-i}) u_i^2(a_i,\boldsymbol{a}_{-i}) - \pi_i^2(a_i) u_i^2(a_i,\pi_{-i}) \right) + \mE^2[U_i(a_i)]\\
    \Rightarrow\, \mathrm{Var}(U_i(a_i)) & = M \left(\pi_i(a_i)\sum_{\boldsymbol{a}_{-i}\in \prod_{j\in -i} A_j} \pi_{-i}(\boldsymbol{a}_{-i}) u_i^2(a_i,\boldsymbol{a}_{-i}) - \pi_i^2(a_i) u_i^2(a_i,\pi_{-i}) \right)\\
    & = M \pi_i(a_i) \left(\sum_{\boldsymbol{a}_{-i}\in \prod_{j\in -i} A_j} \pi_{-i}(\boldsymbol{a}_{-i}) u_i^2(a_i,\boldsymbol{a}_{-i}) - \pi_i(a_i) u_i^2(a_i,\pi_{-i}) \right).
\end{align*}

If we increase the number of samples $M$ per iteration as $M_t=\lceil \tfrac{1}{\epsilon_t \gamma^2_t} \rceil$, then also using the fact that $G(\boldsymbol{\pi}) \leq -V(\boldsymbol{\pi})$, we have
\begin{align*}
    \mE[ V(\boldsymbol{\pi}(t+1)) -V(\boldsymbol{\pi}(t)) | \boldsymbol{\pi}(t) ] 
     \leq - \gamma_t V(\boldsymbol{\pi}(t)) + C_4 \gamma_t^2 + C_5 \gamma_t \epsilon_t.
\end{align*}
Taking $\epsilon_t \propto \gamma_t$, and taking expectations, we get
\begin{align*}
    \mE[ V(\boldsymbol{\pi}(t+1)) ]  \leq (1- \gamma_t) \mE[ V(\boldsymbol{\pi}(t)) ]+ C_6 \gamma_t^2.
\end{align*}
Then, by telescoping we have: 
\[
\mE\left[V(\boldsymbol{\pi}(t))\right] \le \Pi_{s=1}^{t-1} (1-\gamma_{s}) V(\boldsymbol{\pi}(0)) + C_6 \sum_{s=1}^{t-1} \gamma^{2}_{s} \Pi_{i=s+1}^{t-1} (1-\gamma_{s})
\]
with $\gamma_{s} = \frac{1}{s+1}$, we get: 
\[
\Pi_{s=1}^{t-1} (1-\gamma_{s}) = \frac{1}{t}
\]
and 
\[
\sum_{s=0}^{t-1} \gamma^{2}_{s} \Pi_{j=s+1}^{t-1}(1-\gamma_{j}) = \sum_{s=0}^{t-1} \frac{1}{(s+1)^{2}} \Pi_{j=s+1}^{t-1} \frac{j}{j+1} \le \frac{\log(t)}{t}.
\]
Therefore, 
\[
\mE[V(\boldsymbol{\pi}(t))]\le \frac{V(\boldsymbol{\pi}(0))}{(t+1)} + \frac{C_{6}\log(t)}{t}.
\]
Using Markov's inequality for the non-negative function $V(\boldsymbol{\pi})$, we conclude that $\lim_{t\rightarrow\infty} V(\boldsymbol{\pi}(t)) =0$ in probability.

\textbf{Note:} With our choice at time-step $t$ of $\gamma_t=\tfrac{1}{t+1}$, $\epsilon_t\propto \gamma_t$ and $M_t = \lceil \tfrac{1}{\epsilon_t \gamma^2_t} \rceil$, we note that the cumulative number of simulator uses until time $T$ is polynomial in $T$ as $\sum_{t=1}^T M_t = \Omega(T^4)$.

We now show that the gap function $V(\boldsymbol{\pi})$ is $C$-smooth. To show this, we show that the gradient of the gap function is Lipschitz continuous. 

\begin{lemma}\label{lem:Csmooth}
$V(\boldsymbol{\pi})$ is $C$-smooth.
\end{lemma}
\begin{proof}
The gradient of $V(\boldsymbol{\pi})$ is given by:
\begin{equation*}
    \nabla V(\boldsymbol{\pi}) =L(\boldsymbol{\pi}) = -F(\boldsymbol{\pi}) + H^T(\boldsymbol{\pi}) ( \boldsymbol{s}^*(\boldsymbol{\pi}, F(\boldsymbol{\pi}))-\boldsymbol{\pi} )  + \boldsymbol{\lambda}(\boldsymbol{\pi}).
\end{equation*}

For any $\boldsymbol{\pi}_1$ and $\boldsymbol{\pi}_2$ we have 
\begin{align*}
\nabla V(\boldsymbol{\pi}_1) - \nabla V(\boldsymbol{\pi}_2) &=  -F(\boldsymbol{\pi}_1) + H^T(\boldsymbol{\pi}_1) ( \boldsymbol{s}^*(\boldsymbol{\pi}_1, F(\boldsymbol{\pi}_1))-\boldsymbol{\pi}_1 )  + \boldsymbol{\lambda}(\boldsymbol{\pi}_1) \\
& \quad +F(\boldsymbol{\pi}_2) - H^T(\boldsymbol{\pi}_2) ( \boldsymbol{s}^*(\boldsymbol{\pi}_2, F(\boldsymbol{\pi}_2))-\boldsymbol{\pi}_2 )  - \boldsymbol{\lambda}(\boldsymbol{\pi}_2)
\end{align*}
By assumption~\ref{assmpt:lipschitz-continuity}, we have that: 
\begin{align*}
\Vert \nabla V(\boldsymbol{\pi}_1) - \nabla V(\boldsymbol{\pi}_2) \Vert_{2} &\le \Vert F(\boldsymbol{\pi}_2) -F(\boldsymbol{\pi}_1) \Vert_{2} + \Vert H^T(\boldsymbol{\pi}_1) \Vert_{2} \Vert \boldsymbol{\pi}_2 - \boldsymbol{\pi}_1 \Vert_{2} \\
& \quad + \Vert H^T(\boldsymbol{\pi}_1) \Vert_{2} \Vert \boldsymbol{s}^*(\boldsymbol{\pi}_1, F(\boldsymbol{\pi}_1)) - \boldsymbol{s}^*(\boldsymbol{\pi}_2, F(\boldsymbol{\pi}_2))\Vert_{2}\\
 &\quad + \Vert H^T(\boldsymbol{\pi}_1) - H^T(\boldsymbol{\pi}_2)\Vert_{2} \Vert \boldsymbol{\pi}_2 - \boldsymbol{s}^*(\boldsymbol{\pi}_2, F(\boldsymbol{\pi}_2)) \Vert_{2}+ \Vert \boldsymbol{\lambda}(\boldsymbol{\pi}_1) - \boldsymbol{\lambda}(\boldsymbol{\pi}_2) \Vert_{2}\\
    &\le \alpha_{V} \Vert \pi_{1} - \pi_{2} \Vert + \alpha_{H}(1+\alpha_{s})\Vert \pi_{1} - \pi_{2} \Vert + 2\alpha_{H}\vert A \vert \Vert \pi_{1} - \pi_{2} \Vert \\
    & \quad + c_0\vert A \vert \Vert \pi_{1} - \pi_{2} \Vert\\
    &= C \Vert \pi_{1} - \pi_{2} \Vert 
\end{align*}
where, $C = \alpha_{V} + \alpha_{H}(1+\alpha_{S})+2\alpha_{H}\vert A \vert+c_0$. It is easily shown by computing the gradient that $\boldsymbol{\lambda}(\boldsymbol{\pi})$ is Lipschitz. The $C$-smoothness of $V(\boldsymbol{\pi})$ allows us to claim the bound in \eqref{eq:ineqVgen}.\end{proof}


\section{Gap Function Analysis}\label{app:gapfunction}
We start by reminding the reader of the gap function from \eqref{eqn:newgapi} and \eqref{eqn:newgap}:
\begin{align*}
\begin{split}
    {V}_i(\boldsymbol{\pi}) & \eqdef \max_{s \in \Delta(A_i)} \langle s - \pi_i, \nabla_{\pi_i} u_i^{f_i}(\boldsymbol{\pi})\rangle - \eta \mathrm{KL}(s,\pi_i) \\
    & = \eta \log\left(\left\langle \pi_i, \exp\left(\frac{1}{\eta}\nabla_{\pi_i} u_i^{f_i}(\boldsymbol{\pi})\right)\right\rangle \right)   - \eta \left\langle \pi_i, \frac{1}{\eta} \nabla_{\pi_i} u_i^{f_i}(\boldsymbol{\pi})\right\rangle, 
\end{split}\\
    \text{and }  {V}(\boldsymbol{\pi}) & \eqdef \sum_ {i\in [N]} \bar{V}_i(\boldsymbol{\pi}).
\end{align*}
We also remind the reader of the VI characterization of a GQRE from Lemma~\ref{lem:verify-gqre-linear}, namely, a strategy profile $\boldsymbol{\pi}^*$ is a GQRE if and only if
\begin{align*}
    \max_{a_i\in A_i} \langle \delta_{a_i} - \pi_i^*, \nabla_{\pi_i} u_i^{f_i}(\pi^*) \rangle \leq 0,~\forall i\in [N].
\end{align*}
Whereas the relationship above is written for pure actions, by standard convex analysis, it extends to the following: a strategy profile $\boldsymbol{\pi}^*$ is a GQRE if and only if
\begin{align}\label{eq:VI-GQRE}
    \max_{s_i\in A_i} \langle s_i - \pi_i^*, \nabla_{\pi_i} u_i^{f_i}(\pi^*) \rangle \leq 0,~\forall i\in [N].
\end{align}

We start by proving that $V(\boldsymbol{\pi}^*)=0$ for a GQRE $\boldsymbol{\pi}^*$. By the properties of the KL divergence function, we have
\begin{align*}
    \langle \pi_i^* - \pi_i^*, \nabla_{\pi_i^*} u_i^{f_i}(\boldsymbol{\pi}^*)\rangle - \eta \mathrm{KL}(\pi^*,\pi_i^*)=0
\end{align*}
Then, by \eqref{eq:VI-GQRE} and the properties of the KL divergence function, for any $\boldsymbol{s}\neq \boldsymbol{\pi}^*$ for all $i\in [N]$, 
\begin{align*}
    \langle s_i - \pi_i, \nabla_{\pi_i} u_i^{f_i}(\boldsymbol{\pi})\rangle \leq 0,
\end{align*}
with the inequality strict for at least one $i\in [N]$. Further, $\mathrm{KL}(s_i,\pi_i^*)\geq 0$ for all $i\in[N]$ and strictly positive for at least one $i\in [N]$. Therefore, we have 
\begin{align*}
    \langle s_i - \pi_i^*, \nabla_{\pi_i} u_i^{f_i}(\boldsymbol{\pi}^*)\rangle - \eta \mathrm{KL}(s_i,\pi_i^*) \leq 0,
\end{align*}
and hence, $V(\boldsymbol{\pi}^*)=0$.

Next, let $\boldsymbol{\pi}$ which not a GQRE, be such that $\pi_i$ has full support for all $i\in[N]$. Then, by \eqref{eq:VI-GQRE}, there is at least one $i\in [N]$, say $1$ after relabeling, where there is an $s_1$ such that
\begin{align*}
    \langle s_1 - \pi_1, \nabla_{\pi_1} u_1^{f_1}(\boldsymbol{\pi})\rangle - \eta \mathrm{KL}(s_1,\pi_1) = \epsilon_1 > 0.
\end{align*}
For $\delta \in (0,1)$, set $\tilde{s}_1(\delta)=\pi_1+\delta (s_1-\pi_1)$, then for $\delta\ll 1$ we have
\begin{align*}
    \mathrm{KL}(\tilde{s}_1,\pi_1) & =  \sum_{a\in A_1} (\pi_{1,a} + \delta (s_{1,a}-\pi_{1,a})) \log\left(1+ \delta \left(\frac{s_{1,a}}{\pi_{1,a}}-1 \right) \right) \\
    & \approx - \delta^2 \sum_{a\in A_1} \frac{(s_{1,a}-\pi_{1,a})^2}{\pi_{1,a}},
\end{align*}
with error terms being higher order in $\delta$, and hence, negligible. Then, we have
\begin{align*}
    \langle \tilde{s}_1(\delta) - \pi_1, \nabla_{\pi_1} u_1^{f_1}(\boldsymbol{\pi})\rangle - \eta \mathrm{KL}(\tilde{s}_1(\delta),\pi_1) & \approx \delta \epsilon - \delta^2 \eta \sum_{a\in A_1} \frac{(s_{1,a}-\pi_{1,a})^2}{\pi_{1,a}} \\
    & = \delta \left(1- \delta  \eta \sum_{a\in A_1} \frac{(s_{1,a}-\pi_{1,a})^2}{\pi_{1,a}}\right).
\end{align*}
Then, we can make $\delta$ small enough that 
$$ \delta  \eta \sum_{a\in A_1} \frac{(s_{1,a}-\pi_{1,a})^2}{\pi_{1,a}} < 1.$$
Hence, $V_1(\boldsymbol{\pi})>0$, and since $V_i(\boldsymbol{\pi})\geq 0$ for all $i\in [N]$ for any $\boldsymbol{\pi}$ (take $\boldsymbol{s}=\boldsymbol{\pi}$ or $s_i=\boldsymbol{\pi}_i$ for all $i\neq 1$), it follows that $V(\boldsymbol{\pi})>0$.

Note that our gap function is $0$ for all pure strategy profiles. However, in Algorithm~\ref{algo:fw-decentralized} we always operate at fully mixed strategies, and there we have negative drift except for the unique fully mixed equilibrium, so the issue of the gap function being zero at pure strategy profiles doesn't cause trouble.

\section{Additional Numerical Simulations}\label{app:add_num_sim}

\textbf{Environment Setup:}
The experiments are done on a MacBook Air with an Apple M1 chip, 16 GB memory and 10 core CPU. All codes are written in Python3 using several open source packages. The running time for all experiments ranges from less than a minute to a few hours. 

\subsection{Additional Numerical Result for Strongly Monotone and Rank-$k$ games}

\begin{figure}[h]
  \centering
    \includegraphics[width=0.8\linewidth]{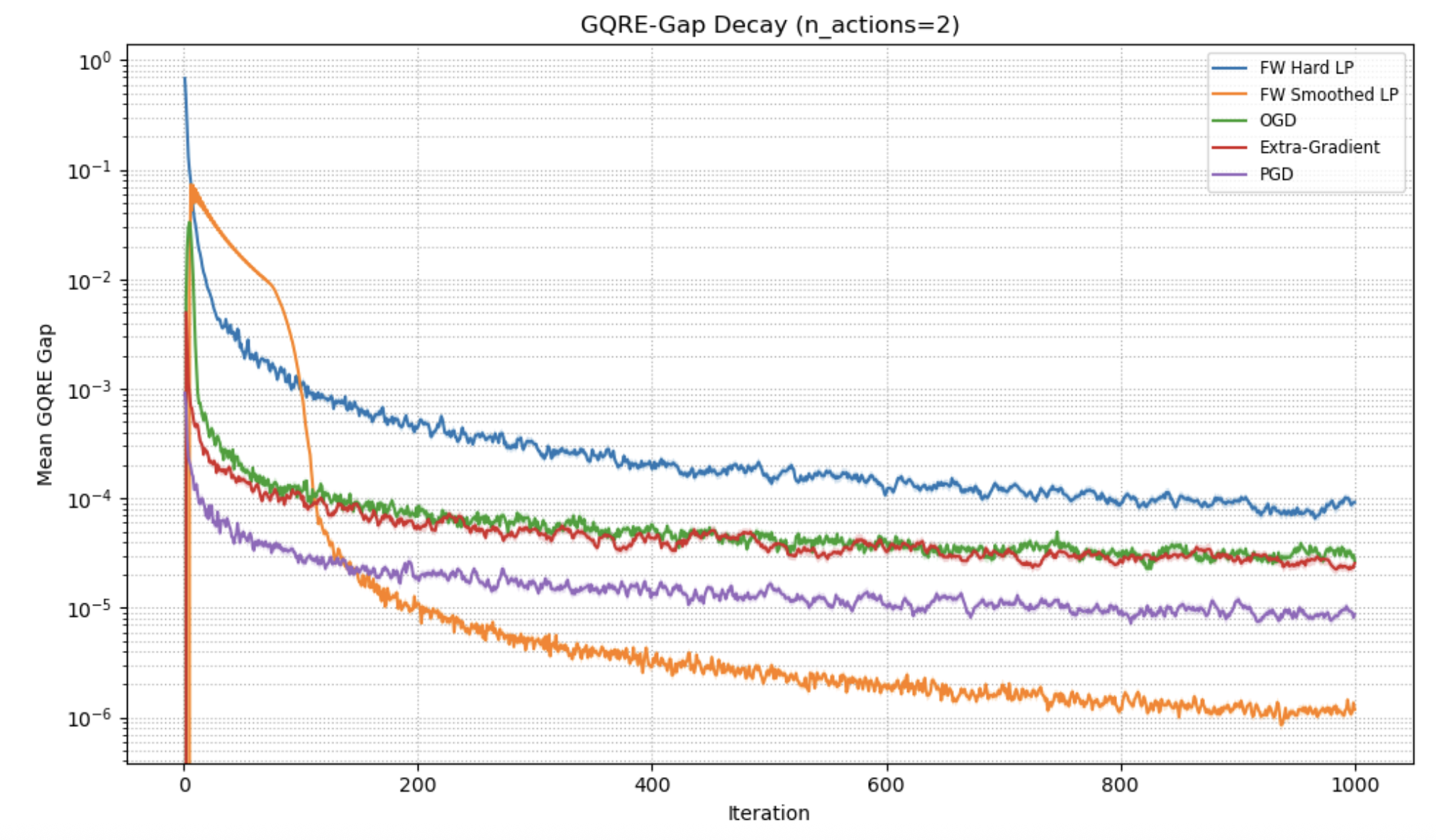}
  \caption{GQRE gap in the matching pennies game.}
  \label{fig:matching_pennies}
\end{figure}

We start by presenting results for 
\textit{Matching Pennies:}
This is a class game with $2\times2$ payoff matrices:
$A = \begin{pmatrix}1 & -1\\-1 & 1\end{pmatrix}$,
and
$B = -A$.
It is a zero-sum game (so rank is 0) in which each player tries to match/mismatch the other's action with a unique mixed-strategy NE. 
Figure~\ref{fig:matching_pennies} presents the Mean GQRE Gap, i.e., the gap from the GQRE for Algorithm \ref{algo:fw-decentralized} (labelled FW Smoothed LP), and we see that our algorithm achieves a very low gap in about 1000 iterations, an order of magnitude better than the next best.

\begin{figure}[h]
    \centering

        \centering
        \includegraphics[width=0.495\linewidth]{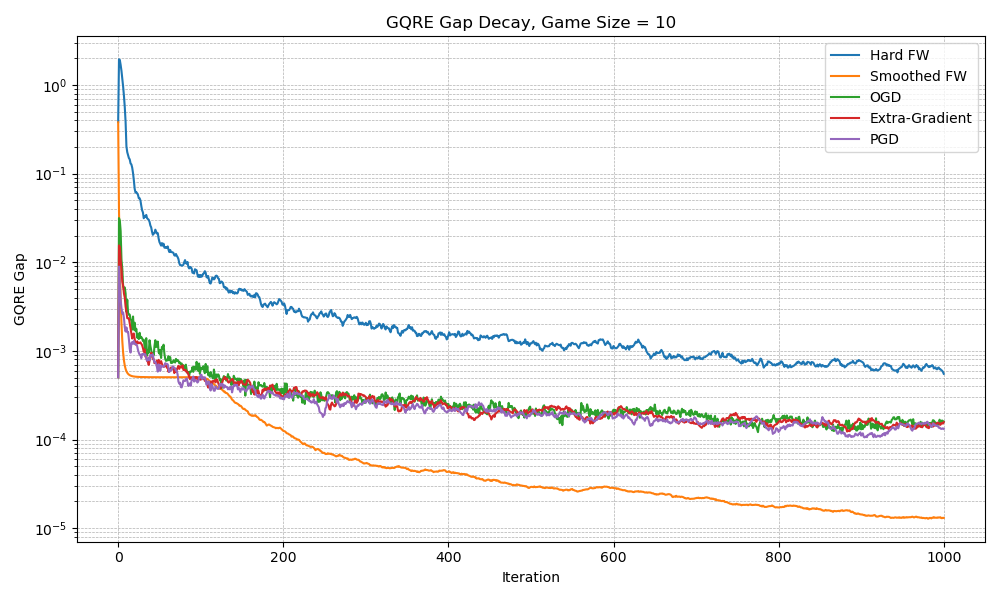}
        \includegraphics[width=0.495\linewidth]{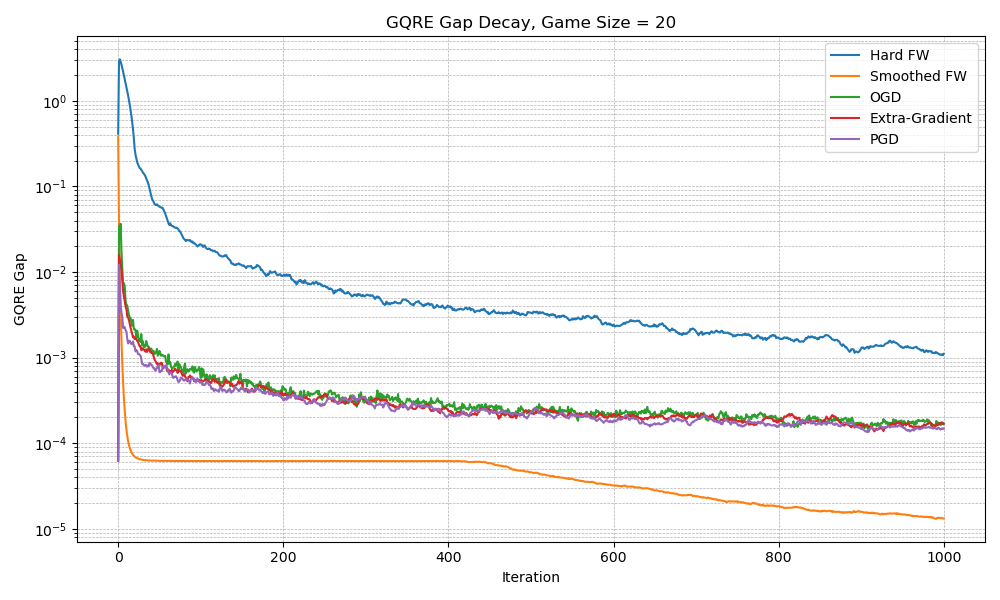}

    \caption{GQRE gap decay for strongly monotone games with sizes $n = 10, 20$.}
    \label{fig:strongly_monotone_small}
\end{figure}

\begin{figure}[h]
  \centering

  \begin{minipage}[t]{0.48\linewidth}
    \centering
    \includegraphics[width=\linewidth]{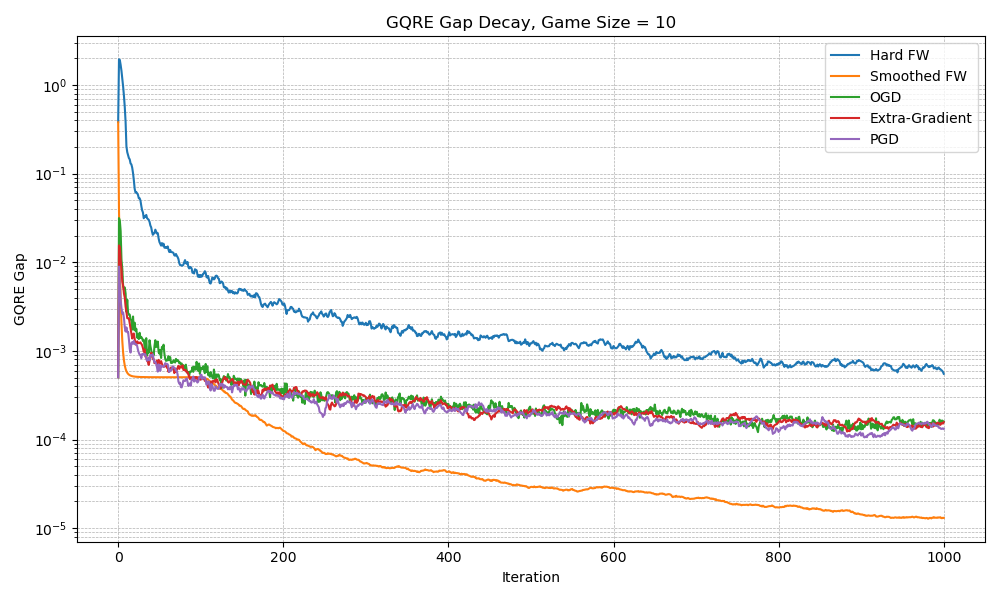}
  \end{minipage}
  \hfill
  \begin{minipage}[t]{0.48\linewidth}
    \centering
    \includegraphics[width=\linewidth]{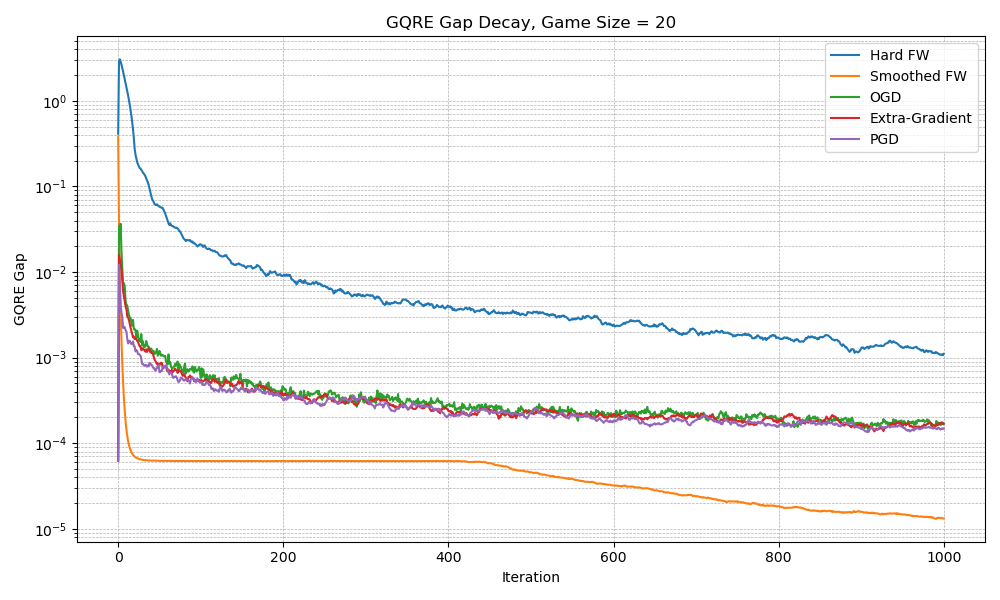}
  \end{minipage}

  \begin{minipage}[t]{0.48\linewidth}
    \centering
    \includegraphics[width=\linewidth]{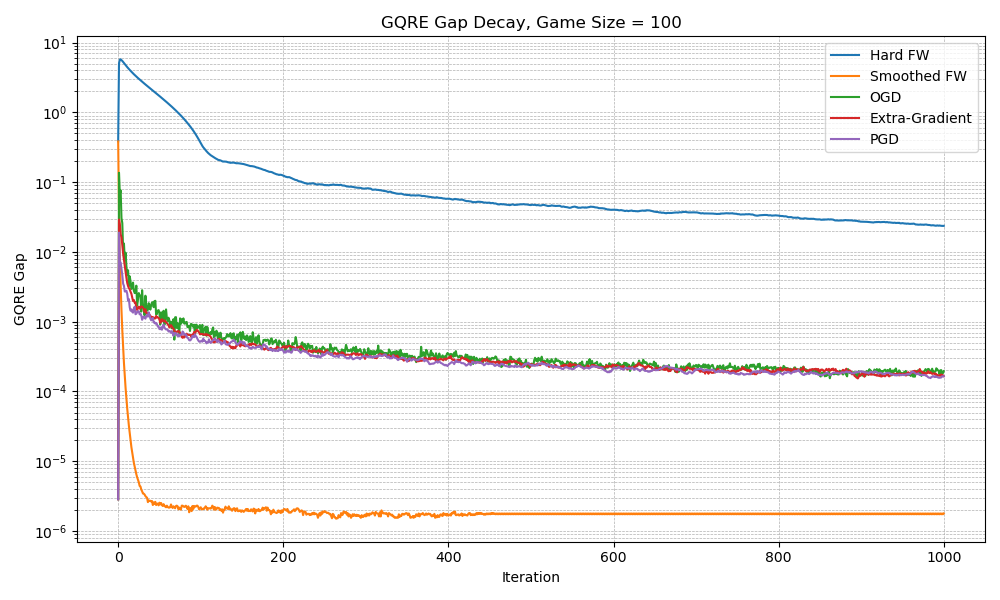}
  \end{minipage}
  \hfill
  \begin{minipage}[t]{0.48\linewidth}
    \centering
    \includegraphics[width=\linewidth]{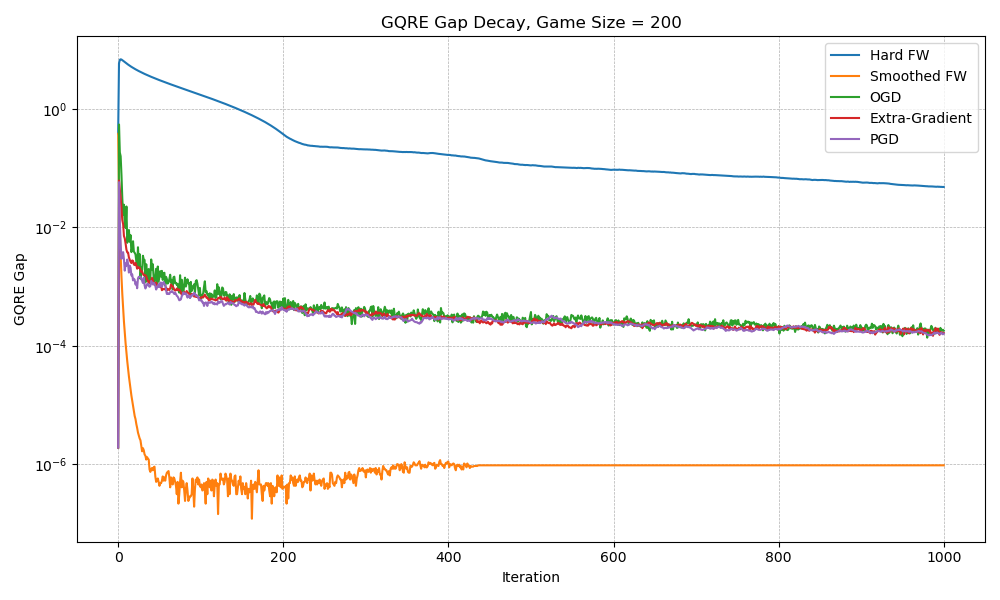}
  \end{minipage}

  \caption{GQRE gap decay on strongly monotone games of increasing size. Smoothed methods consistently outperform in both convergence rate and stability across larger games.}
  \label{fig:stronglymono_grid}
\end{figure}

\begin{figure}[h]
  \centering

  \begin{minipage}[t]{0.48\linewidth}
    \centering
    \includegraphics[width=\linewidth]{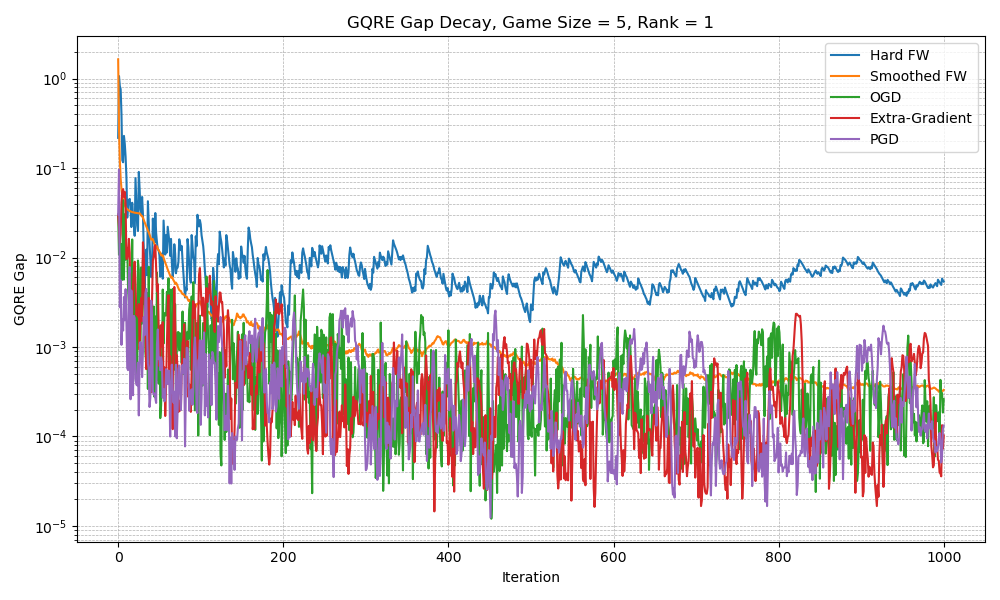}
  \end{minipage}
  \hfill
  \begin{minipage}[t]{0.48\linewidth}
    \centering
    \includegraphics[width=\linewidth]{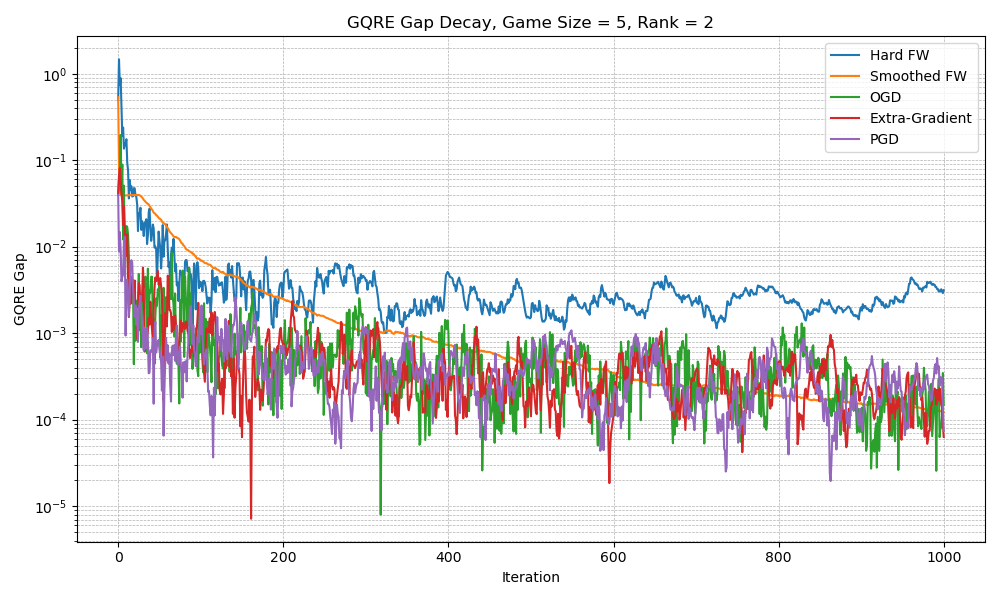}
  \end{minipage}

  \begin{minipage}[t]{0.48\linewidth}
    \centering
    \includegraphics[width=\linewidth]{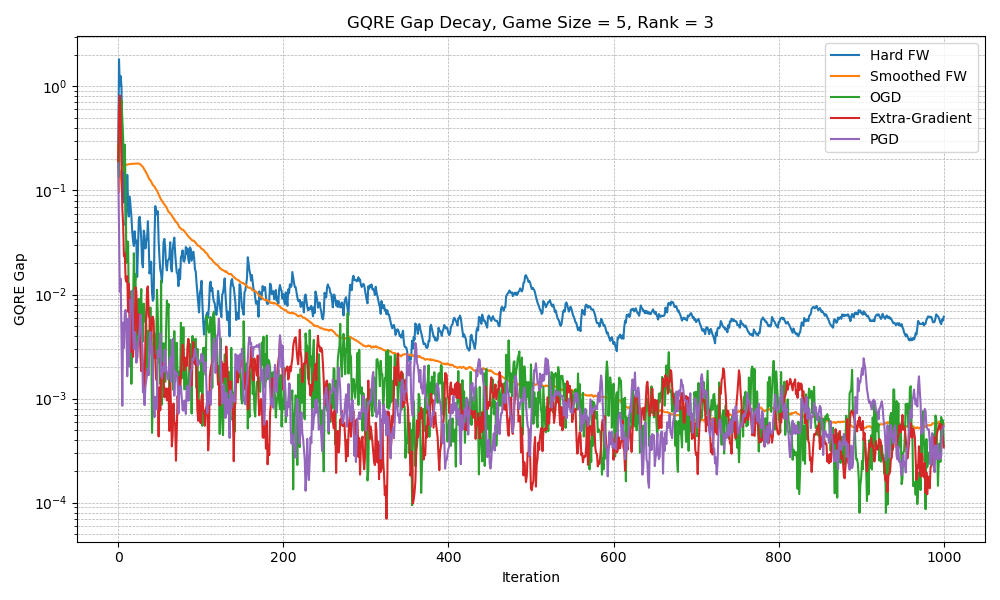}
  \end{minipage}
  \hfill
  \begin{minipage}[t]{0.48\linewidth}
    \centering
    \includegraphics[width=\linewidth]{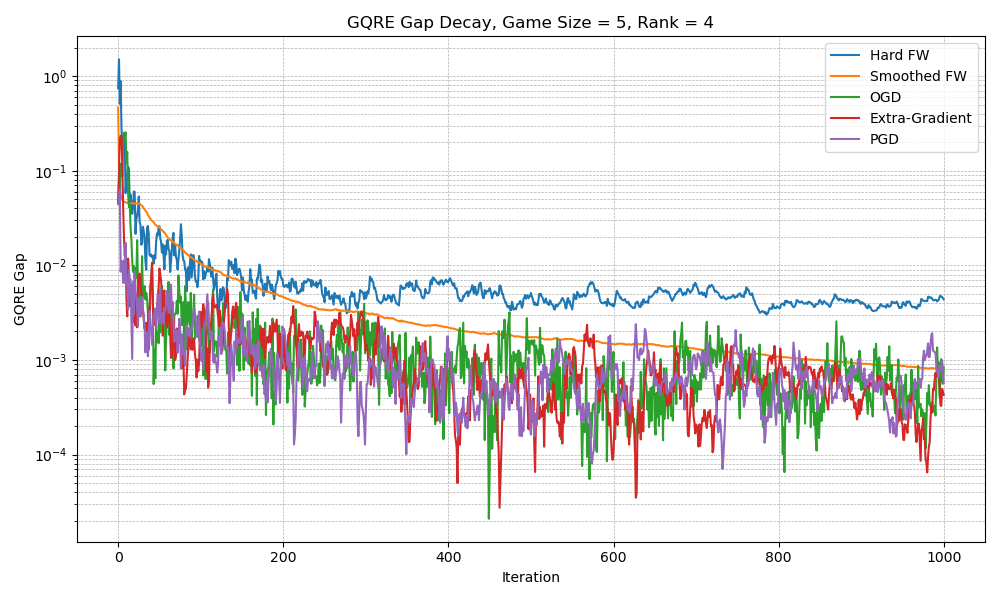}
  \end{minipage}

  \caption{GQRE gap decay for games with size 5 and different ranks.}
  \label{fig:gqre_gap_all}
\end{figure}

Additional results are reported for strongly-monotone games in Figures~\ref{fig:strongly_monotone_small} and ~\ref{fig:stronglymono_grid}. Again, the performance of Algorithm~\ref{algo:fw-decentralized} stands out. Similarly, in Figure~\ref{fig:gqre_gap_all} we present the peformance of Algorithm~\ref{algo:fw-decentralized} relative to the baselines for games of different ranks. Here, Algorithm~\ref{algo:fw-decentralized} show comparable performance but much better stability.

\begin{figure}[h]
  \centering

  \begin{minipage}[t]{0.48\linewidth}
    \centering
    \includegraphics[width=\linewidth]{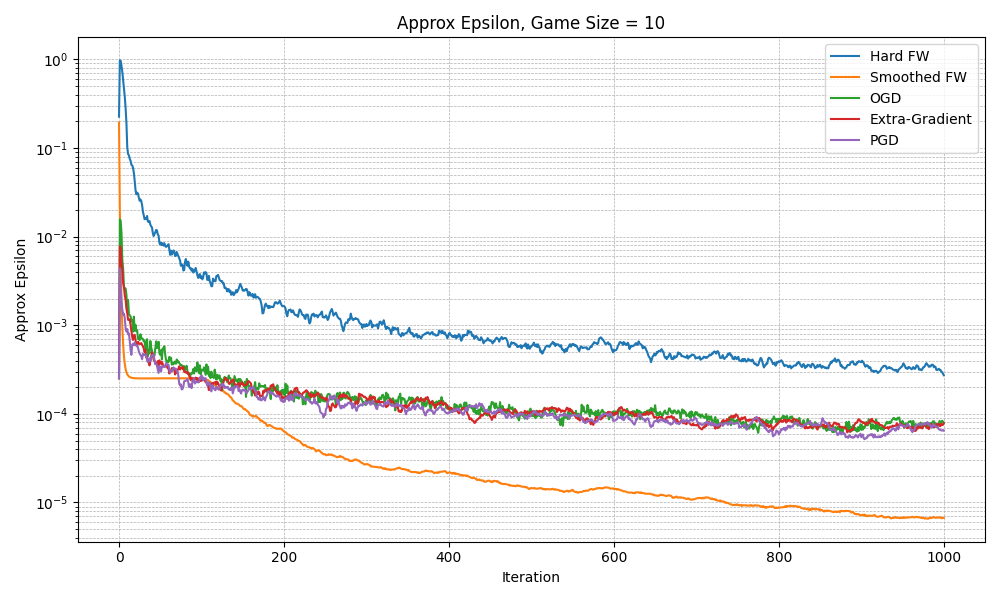}
  \end{minipage}
  \hfill
  \begin{minipage}[t]{0.48\linewidth}
    \centering
    \includegraphics[width=\linewidth]{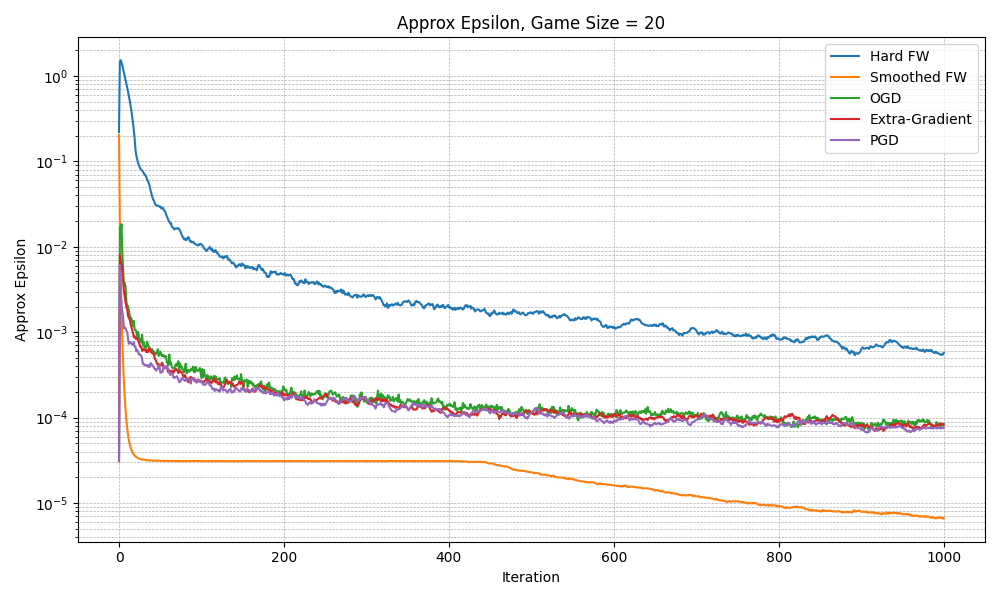}
  \end{minipage}

  \begin{minipage}[t]{0.48\linewidth}
    \centering
    \includegraphics[width=\linewidth]{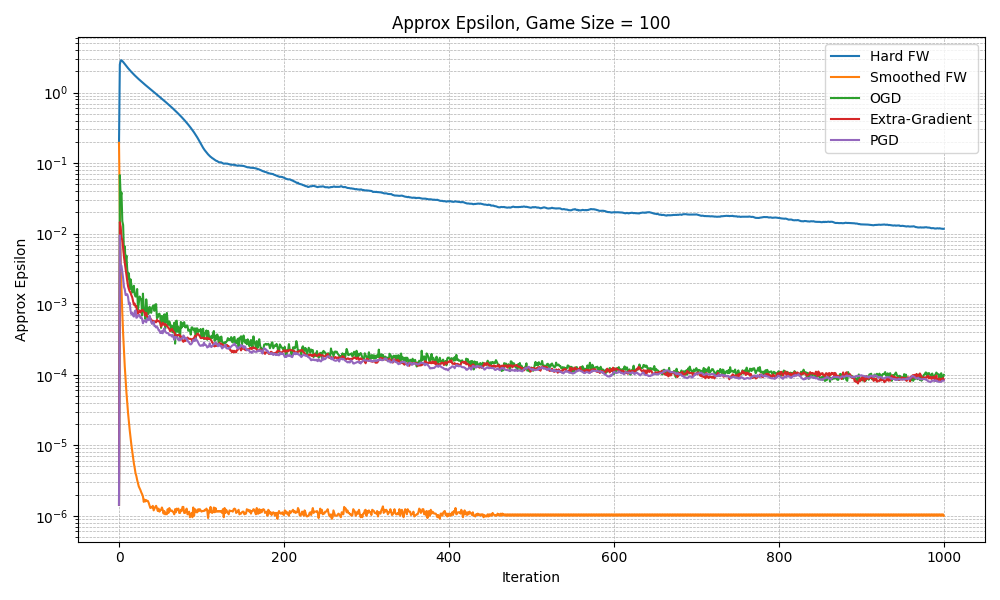}
  \end{minipage}
  \hfill
  \begin{minipage}[t]{0.48\linewidth}
    \centering
    \includegraphics[width=\linewidth]{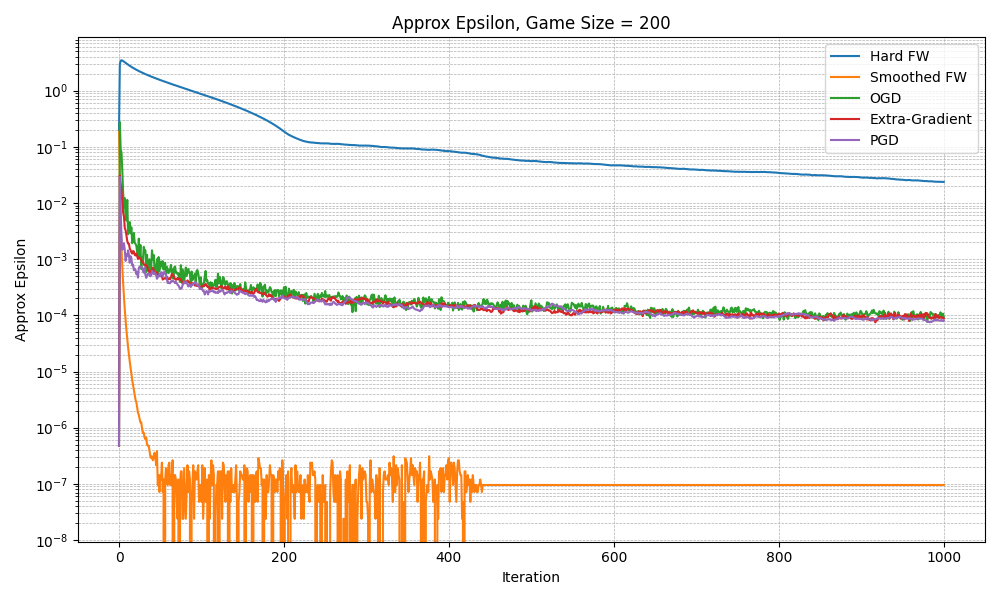}
  \end{minipage}

  \caption{Approx $\epsilon$-GQRE vs iteration for strongly monotone games.}
  \label{fig:approx_epsilon_gap}
\end{figure}








\subsection{Additional Numerical Results with Uniform-Mix Frank--Wolfe}

\begin{figure}[h!]
  \centering
  \begin{subfigure}[t]{0.7\linewidth}
    \centering
    \includegraphics[width=\linewidth]{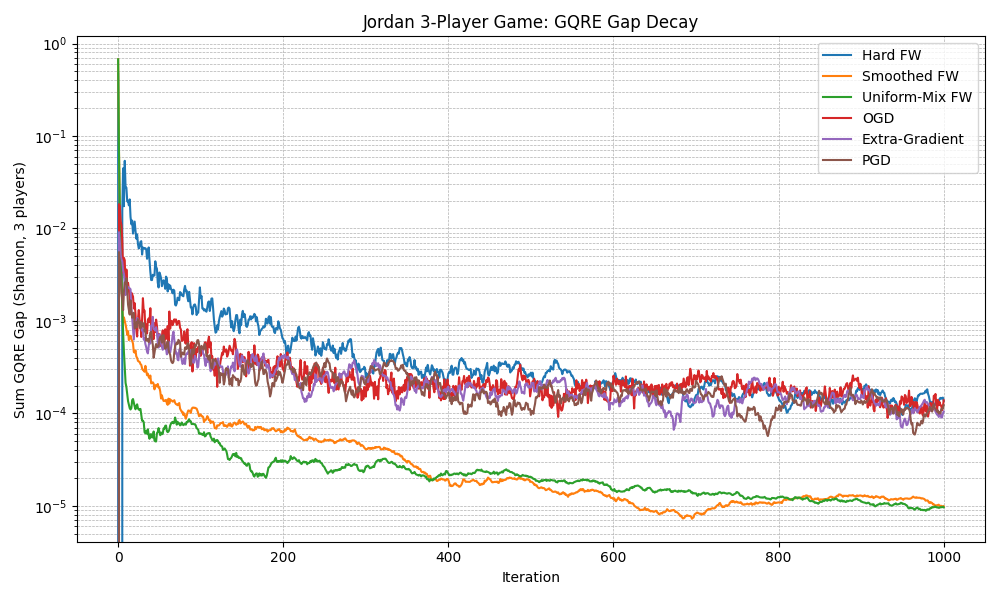}
    \caption{Jordan’s 3-player game.}
  \end{subfigure}

  \caption{Sum GQRE gap decay in Jordan’s 3-player cyclic game. 
  Uniform-Mix FW closely tracks Smoothed FW, converging to near-zero gaps in a non-monotone multi-agent setting.}
  \label{fig:jordan}
\end{figure}

We evaluate the Uniform-Mix Frank--Wolfe (FW) update against Algorithm~\ref{algo:fw-decentralized} and other baselines on Jordan's three-player, two-action game \citep{jordan1993three} in  Figure~\ref{fig:jordan}.



\subsection{Jordan’s 3-Player Game: Stability vs.\ \texorpdfstring{$\lambda$}{lambda} and Sampling \texorpdfstring{$M$}{M}}

Again, we study Jordan’s three-player, two-action game from \citep{jordan1993three}, in which the unique symmetric QRE is $(0.5,0.5,0.5)$. 
For the smoothed Frank--Wolfe (FW) update, Assumption~2.2 (ensuring uniqueness and local stability) is satisfied for small $\lambda$ (here $\lambda<1$). 
For larger $\lambda$, uniqueness may fail and equilibria can become unstable, consistent with the instability phenomena relative to uncoupled dynamics~\citep{hart2003uncoupled}. 

\begin{figure}[h!]
  \centering
  \includegraphics[width=0.9\linewidth]{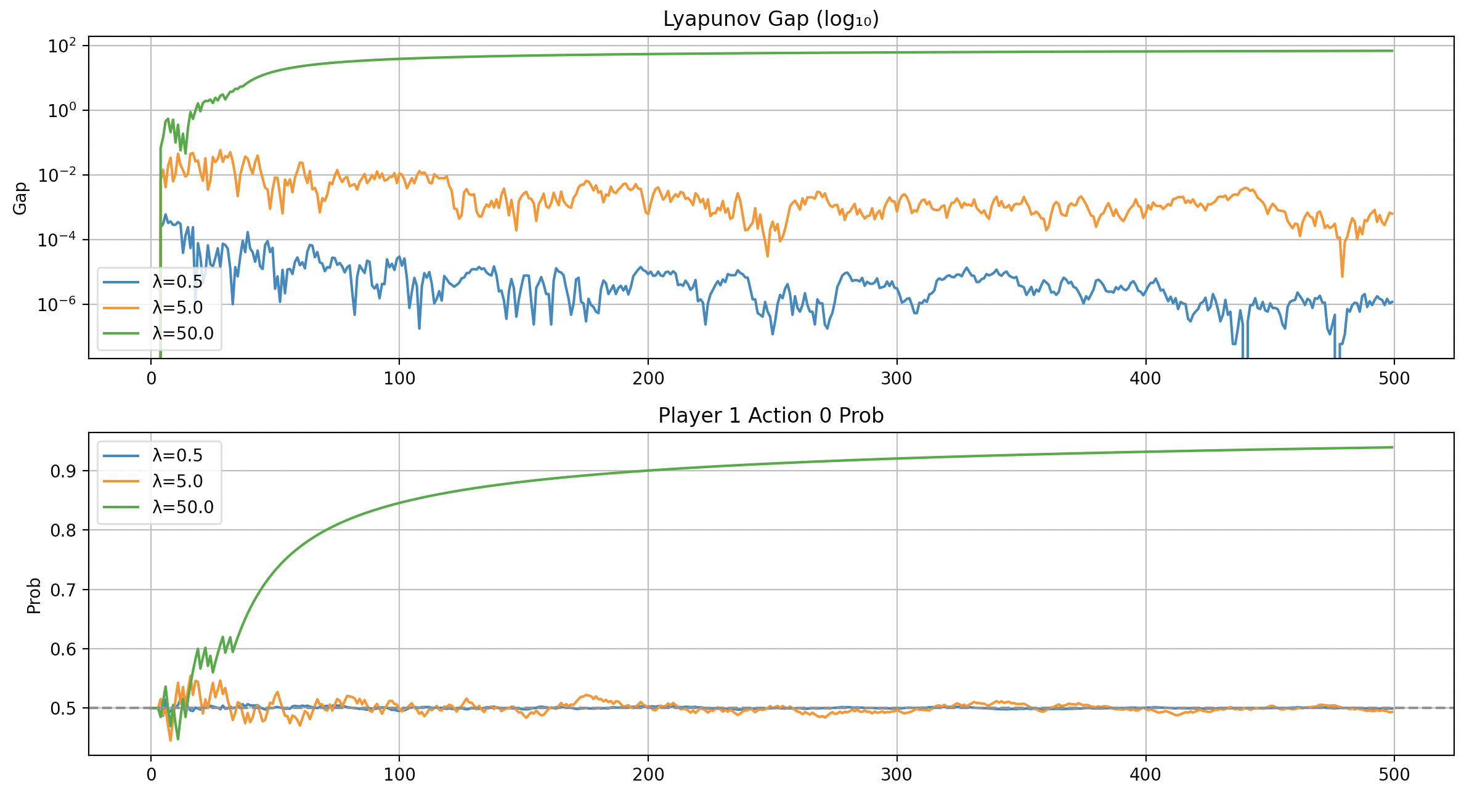}
  \caption{Jordan’s three-player, two-action game with $\lambda \in \{0.5,5,50\}$. 
  Top: Lyapunov gap (log$_{10}$). 
  Bottom: Player~1’s action-$0$ probability. 
  Small $\lambda$ yields robust convergence, moderate $\lambda$ produces mixed outcomes, and large $\lambda$ leads to instability.}
  \label{fig:jordan-lyapunov}
\end{figure}

Figure~\ref{fig:jordan-lyapunov} shows simulation results for $\lambda \in \{0.5,5,50\}$. 
The top panel plots the Lyapunov gap (log scale), while the bottom panel tracks Player~1’s probability of playing action~0. 
For $\lambda=0.5$, the gap decays rapidly and the strategy converges toward the symmetric GQRE. 
At $\lambda=5$, convergence occurs more slowly and is sensitive to sampling noise. 
At $\lambda=50$, the dynamics fail to converge and drift toward the boundary of the simplex, illustrating instability.

\paragraph{Convergence vs.\ sampling $M$.}
We conducted 100 independent experiments for each $(\lambda,M)$ and measured the proportion with final gap $<10^{-3}$, and report the results in 
Table~\ref{tab:jordan-convergence}. 
When $\lambda=0.5$, convergence is robust for all $M$. 
For $\lambda=5$, larger $M$ substantially improves convergence, while for $\lambda=50$ convergence never occurs, even at high $M$. 

\begin{table}[h!]
\centering
\begin{tabular}{lccc}
\toprule
$M$ & $\lambda=0.5$ & $\lambda=5$ & $\lambda=50$ \\
\midrule
$500$  & $100.0\%$ & $37.0\%$ & $0.0\%$ \\
$1000$ & $100.0\%$ & $58.0\%$ & $0.0\%$ \\
$5000$ & $100.0\%$ & $98.0\%$ & $0.0\%$ \\
\bottomrule
\end{tabular}
\caption{Convergence rates (final gap $<10^{-3}$) across $\lambda$ and $M$ in Jordan’s game.}
\label{tab:jordan-convergence}
\end{table}

These results confirm the stability picture: when Assumption~2.2 holds (small $\lambda$), convergence is guaranteed; in the intermediate regime (moderate $\lambda$), variance reduction via larger $M$ can restore convergence; for large $\lambda$, equilibria are unstable and convergence is not possible, in line with the impossibility results for uncoupled dynamics~\cite{hart2003uncoupled}.

\subsection{Heterogeneous Regularizers: Mixed $f_i$ Functions Across Players}

To understand the potential benefit of assigning 
different $f_i$ functions to different agents, we ran simulations with heterogeneous 
regularizers. Specifically, for the rank-$k$ game of size $5$, we considered all 
pairwise combinations of four divergences: Shannon entropy (KL), Rényi ($\alpha=0.5$), 
$0.5\| \cdot \|_2^2$ relative to uniform, and the total variation (TV) distance to 
uniform. In each run, we fixed $\lambda=1$ and used our smoothed Frank--Wolfe algorithm.



\paragraph{Distances from uniform.}
We also computed the KL, L2, and TV distances of each resulting distribution 
from the uniform distribution. Results (Table~\ref{tab:rankk-distances}) show 
that entropy-based equilibria remain closest to uniform, Rényi pushes farthest 
away, and TV regularization produces distributions almost indistinguishable 
from uniform.

\begin{table}[h!]
\centering
\begin{tabular}{lccc}
\toprule
$f_1$--$f_2$ & KL divergence & L2 distance & TV distance \\
\midrule
Entropy--Entropy & 0.0319 & 0.1180 & 0.1022 \\
Entropy--TV      & 0.0119 & 0.0712 & 0.0620 \\
L2--Entropy      & 0.0751 & 0.1855 & 0.1631 \\
L2--TV           & 0.0333 & 0.1213 & 0.1067 \\
Rényi--Rényi     & 0.1992 & 0.3156 & 0.2822 \\
TV--TV           & 0.0000 & 0.0008 & 0.0008 \\
\bottomrule
\end{tabular}
\caption{Distances of Player~1 distributions from uniform under 
different regularizer pairs. Entropy and TV combinations yield 
the closest-to-uniform distributions, while Rényi divergence pushes farthest apart.}
\label{tab:rankk-distances}
\end{table}

\paragraph{Convergence surrogate.}
To evaluate convergence, we use the gap value at $T=1000$ iterations as a 
surrogate measure. Table~\ref{tab:rankk-gap} shows these values. Mixed 
regularizers yield notable improvements: Entropy--TV and L2--Entropy 
significantly reduce the final gap compared to their homogeneous counterparts. 
In contrast, homogeneous cases of Rényi and TV divergences do not benefit from mixing.

\begin{table}[h!]
\centering
\begin{tabular}{lcccc}
\toprule
 & Entropy & L2 & Rényi & TV \\
\midrule
Entropy & 0.000081 & 0.054651 & 0.603757 & 0.049424 \\
L2      & 0.052204 & 0.078403 & 0.622428 & 0.098707 \\
Rényi   & 0.313565 & 0.276359 & 0.967499 & 0.813753 \\
TV      & 0.039242 & 0.136138 & 0.855256 & 0.039518 \\
\bottomrule
\end{tabular}
\caption{Surrogate convergence: gap at $T=1000$ for each pair $(f_1,f_2)$. 
Mixed combinations (e.g., Entropy--TV, L2--Entropy, L2--TV) improve 
convergence rates relative to homogeneous choices.}
\label{tab:rankk-gap}
\end{table}

\paragraph{Summary.}
These results show that heterogeneous choice of regularizers can 
be beneficial. In particular, Entropy--TV and L2--based mixtures 
consistently improve convergence compared to homogeneous choices. 
This provides a concrete example where using different $f_i$ across 
players yields different GQRE and also enhances the robustness of the algorithm.

\section{Examples of Penalty Functions and Responses}\label{app:example_responses}

\subsection{Fair Quantal Response} 
We aim to solve the optimization problem:
\begin{equation*}
\max_{\mathbf{p}} \quad \sum_{a \in A} p(a) \lambda u(a) - \frac{1}{2} \sum_{a \in A} \left| p(a) - \frac{1}{|A|} \right|\\
\text{s.t.}\ \sum_{a \in A} p(a) = 1, \quad p(a) \geq 0 \quad \forall a \in A,
\end{equation*}
which we rewrite through scaling as
\begin{align*}
    \max_{\mathbf{p}} \quad \sum_{a \in A} p(a) 2\lambda u(a) - \sum_{a \in A} \left| p(a) - \frac{1}{|A|} \right|\\
\text{s.t.}\ \sum_{a \in A} p(a) = 1, \quad p(a) \geq 0 \quad \forall a \in A.
\end{align*}
Let \( n = |A| \), so the uniform distribution is given by \( p(a)^{\text{uniform}} = \frac{1}{n} \). Introducing the Lagrange multiplier \( \mu \) for the sum probability constraint, the Lagrangian is given by:
\begin{equation*}
\mathcal{L} = \sum_{a \in A} p(a) 2\lambda u(a) - \sum_{a \in A} \left| p(a) - \frac{1}{n} \right|
- \mu \left(\sum_{a \in A} p(a) - 1 \right) 
\end{equation*}
Since we have separability, we can equivalently solve the following piece-wise linear continuous function with a kink at $\frac{1}{\vert A\vert}$ for each action $a\in A$:
\[
\max_{p \in [0,1]} p \left(2\lambda u(a)-\mu\right) - | p -1/n|
\]
In the segment $[0,1/n]$, the function is 
$$-1/n + p \left(2\lambda u(a)-\mu+1\right), $$
and in the segment $[1/n,1]$, the function is
$$ 1/n +  p \left(2\lambda u(a)-\mu-1\right). $$
Hence, we have the maximizer for each $\mu$ given by
\begin{align*}
    p^*(a;\mu) \in 
    \begin{cases}
        \{0\} & \text{if } \mu > 2\lambda u(a)+1 \\
        [0,1/n] & \text{if } \mu = 2\lambda u(a)+1 \\
        \{1/n\} & \text{if } \mu \in \left(2\lambda u(a)-1,  2\lambda u(a)+1\right) \text{ and } 2\lambda u(a)> 1\\
        [1/n,1] & \text{if } \mu = 2\lambda u(a)-1\text{ and } 2\lambda u(a)\ge 1 \\
        \{1\} & \text{if } \mu < 2\lambda u(a)-1 \text{ and } 2\lambda u(a)> 1 \\
        \{1/n\} & \text{if } \mu \in \left[0,  2\lambda u(a)+1\right) \text{ and } 2\lambda u(a)< 1
    \end{cases}
\end{align*}
Thereafter, we choose $\mu$ to be a value such that $p^*(\cdot;\mu)$ vector lies in the probability simplex. For this we can start with $1+ \tfrac{2\max_a u(a)}{\lambda}$ and then decrease $\mu$ until we get a probability vector. At points where we have multiple solutions, a careful selection may be needed.

\subsection{Wasserstein Distance Penalty}

Let \( \mathbf{p}, \mathbf{p}_0 \in \Delta_n \) be discrete probability distributions over a uniform grid of ordered support points \( x_1 < x_2 < \cdots < x_n \in \mathbb{R} \), denoted by the vector \( \mathbf{x} = (x_1, \dots, x_n)^\top \in \mathbb{R}^n \).

We consider the optimization problem:
\begin{equation}
\max_{\mathbf{p} \in \Delta_n} \left\{
\lambda\mathbf{v}^\top \mathbf{p} -  W_2^2(\mathbf{p}, \mathbf{p}_0)
\right\},
\end{equation}
where:
\begin{itemize}
  \item \( \mathbf{v} \in \mathbb{R}^n \) is a utility vector,
  \item \( \lambda > 0 \) is a regularization parameter,
  \item \( W_2^2(\mathbf{p}, \mathbf{p}_0) \) is the squared Wasserstein-2 distance between \( \mathbf{p} \) and \( \mathbf{p}_0 \).
\end{itemize}

This formulation can be generalized when using a Wasserstein $p$-distance as:
\begin{itemize}
  \item \( p \in \Delta_n \) be a probability distribution over \( n \) actions (decision variable),
  \item \( q \in \Delta_n \) be a reference distribution,
  \item \( u \in \mathbb{R}^n \) be the utility vector,
  \item \( d(i, j) \) denote the ground distance between actions \( i \) and \( j \),
  \item \( \lambda > 0 \) be the regularization parameter.
\end{itemize}

\subsection{Squared Distance Penalty}

The optimization problem here is given by:
\begin{equation}
\max_{\bm{p} \in \Delta_n} \left\{ \lambda\bm{v}^\top \bm{p} -  \left( \bm{x}^\top (\bm{p} - \bm{p}_0) \right)^2 \right\},
\end{equation}
where:
\begin{itemize}
  \item \( \bm{p} \in \Delta_n \) is a probability vector (i.e., \( \bm{p} \geq 0, \sum_i p_i = 1 \)),
  \item \( \bm{v} \in \mathbb{R}^n \) is a utility vector,
  \item \( \bm{x} \in \mathbb{R}^n \) is a vector of support points (e.g., \( \bm{x} = (1, 2, \dots, n)^\top \)),
  \item \( \bm{p}_0 \in \Delta_n \) is a reference distribution,
  \item \( \lambda > 0 \) is a regularization parameter.
\end{itemize}

Let us define:
\[
a := \bm{x}, \qquad b := \bm{x}^\top \bm{p}_0,
\]
then the objective becomes:
\begin{equation}
\min_{\bm{p} \in \Delta_n} \left\{ -\lambda\bm{v}^\top \bm{p} +  (a^\top \bm{p} - b)^2 \right\}.
\end{equation}

Expanding the square term:
\[
(a^\top \bm{p} - b)^2 = \bm{p}^\top aa^\top \bm{p} - 2b a^\top \bm{p} + b^2.
\]

Therefore, the optimization problem becomes:
\begin{equation}
\min_{\bm{p} \in \Delta_n} \left\{ \bm{p}^\top Q \bm{p} + \bm{c}^\top \bm{p} \right\} + \text{const},
\end{equation}
where:
\[
Q =  aa^\top, \qquad \bm{c} = -\lambda\bm{v} - 2 b a.
\]

This is a quadratic program (QP) over the simplex:
\begin{align}
\text{minimize} \quad & \bm{p}^\top Q \bm{p} + \bm{c}^\top \bm{p}, \\
\text{subject to} \quad & \bm{p} \geq 0, \quad \bm{1}^\top \bm{p} = 1.
\end{align}

\paragraph{Application to approximation of Wasserstein-2 Distance in 1D:}
In 1D, with ordered supports and under some assumptions to be detailed below, the Wasserstein-2 distance can be approximated by the squared difference in expected values:
\begin{equation}
W_2^2(\mathbf{p}, \mathbf{p}_0) \approx \left( \mathbb{E}_{\mathbf{p}}[\mathbf{x}] - \mathbb{E}_{\mathbf{p}_0}[\mathbf{x}] \right)^2
= \left( \mathbf{x}^\top (\mathbf{p} - \mathbf{p}_0) \right)^2.
\end{equation}
The approximation can be shown as follows: 
Let \( \mathbf{p} \) and \( \mathbf{p}_0 \) be probability measures on \( \mathbb{R} \) with finite second moments, i.e., \( p, p_0 \in \mathcal{P}_2(\mathbb{R}) \). Denote their cumulative distribution functions by \( F_p \), \( F_{p_0} \), and their quantile functions by \( F_p^{-1} \), \( F_{p_0}^{-1} \). Define:
\[
X := F_{p}^{-1}(U), \quad Y := F_{p_0}^{-1}(U), \quad U \sim \text{Unif}[0,1].
\]
Then the squared 2-Wasserstein distance is given by:
\[
W_2^2(p, p_0) = \int_0^1 \left( F_{p}^{-1}(u) - F_{p_0}^{-1}(u) \right)^2 \, du = \mathbb{E}[(X - Y)^2].
\]
We apply the identity:
\[
\mathbb{E}[(X - Y)^2] = \left( \mathbb{E}[X - Y] \right)^2 + \operatorname{Var}(X - Y),
\]
which yields:
\[
W_2^2(p, p_0) = (\mu_1 - \nu_1)^2 + \operatorname{Var}(X - Y),
\]
where \( \mu_1 := \mathbb{E}[X] \), \( \nu_1 := \mathbb{E}[Y] \).
The approximation $W_2^2(p, p_0) \approx (\mu_1 - \nu_1)^2$ is valid when $
\operatorname{Var}(X - Y) \ll (\mu_1 - \nu_1)^2$, i.e., the variance term is negligible compared to the squared mean difference. This holds when:
\begin{itemize}
\item Translation of means:  

If \( \nu = \mu(\cdot - \delta) \), then \( F_\nu^{-1}(u) = F_\mu^{-1}(u) + \delta \), so
\[
X - Y = -\delta \quad \text{a.s.}, \quad \operatorname{Var}(X - Y) = 0.
\]
Thus:
\[
W_2^2(p,p_0) = \delta^2 = (\mu_1 - \nu_1)^2.
\]
\item Small Perturbations: 

If \( F_\mu^{-1}(u) - F_\nu^{-1}(u) = \delta(u) \) with \( \delta(u) \) small and smooth, then
\[
X - Y = \delta(U), \quad \operatorname{Var}(X - Y) = \operatorname{Var}(\delta(U)) \text{ is small}.
\]
Hence,
\[
W_2^2(p,p_0) = (\mu_1 - \nu_1)^2 + o(1).
\]
\end{itemize}

\subsection{Renyi Divergence Penalty}

We aim to maximize an expected utility function while keeping the probability distribution \( P \) close to a uniform reference distribution \( Q \) using a Rényi divergence penalty. Let:
\begin{itemize}
    \item \( X \) be a discrete random variable with support \( \mathcal{X} = \{x_1, x_2, ..., x_n\} \).
    \item \( P = (p_1, p_2, ..., p_n) \) be the probability distribution over \( X \).
    \item \( Q = (q_1, q_2, ..., q_n) \) be the reference uniform distribution, i.e., \( q_i = \frac{1}{n} \) for all \( i \).
    \item \( U(x) \) be the utility function associated with choosing \( x \).
    \item For \( \alpha \in (0,1), D_\alpha(P \| Q) \) be the Rényi divergence penalty. 
\end{itemize}

The Rényi divergence of order \( \alpha \) between \( P \) and uniform \( Q \) is:
\[
D_\alpha(P \| Q) = \frac{1}{\alpha - 1} \log \sum_{i=1}^{n} p_i^\alpha q_i^{1-\alpha}.
\]
Since \( q_i = \frac{1}{n} \), we substitute:
\[
D_\alpha(P \| Q) = \frac{1}{\alpha - 1} \log \left( \frac{1}{n^{1-\alpha}} \sum_{i=1}^{n} p_i^\alpha \right).
\]

The optimization problem is given by:
\[
\max_{\boldsymbol{p}} \sum_{i=1}^{n} p_i \lambda U(x_i) - D_\alpha(\boldsymbol{p} \| Q)
\]
subject to:
\[
\boldsymbol{p}=(p_1,\dotsc,p_n) \text{ s.t. }\sum_{i=1}^{n} p_i = 1, \quad p_i \geq 0, \quad \forall i.
\]
where \( \lambda >0 \) controls the strength of the penalty. The Lagrangian for $\alpha\in(0,1)$ ($\alpha=1$ case is QR) is
\[
\mathcal{L}(P, \mu) = \sum_{i=1}^{n} p_i \lambda U(x_i) + \frac{1}{1-\alpha} \log \left( \frac{1}{n^{1-\alpha}} \sum_{i=1}^{n} p_i^\alpha \right) + \mu \left( 1- \sum_{i=1}^{n} p_i \right).
\]
Then, taking a derivative with respect to $p_i$ we get
\[
\lambda U(x_i) + \frac{\alpha}{1-\alpha} \frac{p_i^{\alpha - 1}}{\sum_{j=1}^{n} p_j^\alpha} - \mu \leq 0,
\]
with equality if $p_i>0$. From this it follows that
\[
p_i^{\alpha-1} \leq \frac{1-\alpha}{\alpha} (\sum_{j=1}^{n} p_j^\alpha) (\mu-\lambda U(x_i)). 
\]
From the above it follows that the Lagrange multiplier $\mu > \lambda \max_{i} U(x_i) $ and none of the optimal probability values are $0$, so that
\begin{align*}
p_i^{\alpha-1} & = \frac{1-\alpha}{\alpha} (\sum_{j=1}^{n} p_j^\alpha) (\mu-\lambda U(x_i))\\
\Rightarrow\ p_i & \propto (\mu-\lambda U(x_i))^{\frac{-1}{1-\alpha}}.
\end{align*}
Hence, we have
\begin{align*}
    p_i(\mu) = \frac{(\mu-\lambda U(x_i))^{\frac{-1}{1-\alpha}}}{\sum_{j}(\mu-\lambda U(x_j))^{\frac{-1}{1-\alpha}}}.
\end{align*}
Since we also have
\begin{align*}
    \frac{p_i^\alpha}{\sum_{j=1}^{n} p_j^\alpha}& =\frac{1-\alpha}{\alpha} p_i (\mu-\lambda U(x_i)) \\
    \Rightarrow\ 1 & =\frac{1-\alpha}{\alpha} \sum_i p_i(\mu) (\mu-\lambda U(x_i)),
\end{align*}
The correct value of $\mu$ is determined by solving for the unique root of
\begin{align*}
    \frac{\sum_j (\mu-\lambda U(x_j))^{\frac{-\alpha}{1-\alpha}}}{\sum_j (\mu-\lambda U(x_j))^{\frac{-1}{1-\alpha}} } = \frac{\alpha}{1-\alpha}
\end{align*}
with $\mu > \lambda \max_{i} U(x_i)$.

\subsection{Hellinger Distance Penalty}

We consider a Hellinger distance perturbation with uniform distribution as the reference function. Define:
\begin{itemize}
    \item \( p \in \Delta_n \): decision distribution over \( n \) actions,
    \item \( u \in \mathbb{R}^n \): utility vector,
    \item \( q = \left( \frac{1}{n}, \dots, \frac{1}{n} \right) \in \Delta_n \): uniform reference distribution,
    \item \( \lambda > 0 \): regularization parameter.
\end{itemize}
The squared Hellinger distance is defined as:
\[
H^2(p, q) = 1 - \sum_{i=1}^n \sqrt{p_i q_i} = 1 - \frac{1}{\sqrt{n}} \sum_{i=1}^n \sqrt{p_i}
\]




The optimization problem is given by
\begin{align*}
    \max_{\boldsymbol{p}\in \Delta(A)}  \sum_{i=1}^n p_i \lambda u_i + \sqrt{\frac{p_i}{n}}
\end{align*}
The Lagrangian obtained by relaxing the sum constraint is 
\begin{align*}
    L(\boldsymbol{p},\mu) = \sum_{i=1}^n p_i (\lambda u_i-\mu) + \sqrt{\frac{p_i}{n}} + \mu
\end{align*}
Taking derivatives in $p_i$, we get
\begin{align*}
    \lambda u_i-\mu + \frac{1}{2\sqrt{p_i n}} & \leq 0\\
    \Rightarrow \frac{1}{\sqrt{p_i}} & \leq 2\sqrt{n} (\mu-\lambda u_i).
\end{align*}
Hence, we have the following: at the optimum $p_i>0$ for all $i$; and 
\begin{align*}
    p_i = \frac{1}{4 n (\mu - \lambda u_i)^2}
\end{align*}
where $\mu$ is the unique root of
\begin{align*}
    \sum_i \frac{1}{(\mu - \lambda u_i)^2} = 4 n
\end{align*}
with $\mu > \lambda \max_i u_i$.


\section{Proof of Lemma~\ref{lem:verify-gqre-linear}}
\begin{proof} To prove, we can proceed as follows: \\
Step 1: Fix player $i \in [N]$. From Theorem~\ref{thm:gqre-ne}, checking whether $(\optpol{i},\optpol{-i})$ is GQRE is equivalent to checking if $(\optpol{i},\optpol{-i})$ is a NE of $\perturbgame$. This holds if and only if $\optpol{i} \in \arg\max_{\pol{} \in \Delta(A_{i})} u_i^{f_i}(\pol{},\pi_{-i}^*)$. The latter is equivalent to the following using the VI characterization: 
\begin{align*}
    \optpol{i}\text{ satisfies }\langle \nabla_{\pi_i}u_i^{f_i}(\boldsymbol{\optpol{}}), \pi_i - \optpol{i} \rangle\leq 0,\ \forall \pi_i \in \Delta(A_i).
\end{align*}
The LHS is a linear program, and the check needs to be performed only over the set of pure-strategies since $\Delta(A_{i})$ is a simplex and the maximum occurs at the extreme points.\\
Step 2: This follows immediately from the definition of an $\epsilon$-GQRE.

Note that the utilities being multi-linear in the strategies, determining the value of $\epsilon$ for a given non-NE strategy profile $\boldsymbol{\pi}$ in a finite normal form game only needs evaluation of pure actions. 
\end{proof}

\section{Proof of Proposition~\ref{thm:properties-q-response} and Theorem~\ref{thm:gqre-ne}}

The argument for the proof of Proposition~\ref{thm:properties-q-response} is the following.
\begin{proof}
    For every given $\boldsymbol{\pi}_{-i}$, the optimization function in \eqref{eqn:generalized-q-response} is a convex optimization problem when $f_i$ is a convex function. Then, by general results in convex optimization, an optimizer exists and the GQR is well-defined. If, in addition, $f_i$ is strictly convex, then the objective of the optimization problem is strictly convex which establishes the uniqueness of the optimizer. Finally, using Berge's Theorem of the Maximum, as the objective is jointly continuous in $\boldsymbol{\pi}$ and a unique optimizer exists when $f_i$ is strictly convex, the optimizer is a continuous function of the parameter $\boldsymbol{\pi}_{-i}$.
\end{proof}

The argument for the proof of Theorem~\ref{thm:gqre-ne} is the following.
\begin{proof}
We note that for a strictly convex perturbation $f_i$, the $\cG_f$ is a concave game and therefore, the existence of Nash Equilibrium follows from~\cite{rosen1965existence}. Further, the diagonal domninance criteria used by~\cite{rosen1965existence} follows from Assumption~\ref{assmpt:diagonal-dominance}. 
\end{proof}

\end{document}